\documentstyle[epsf,emulateapj]{article}

\received{17 March 1997}
\accepted{23 December 1997}

\slugcomment{To appear in ApJ., May 10 1998}

\lefthead{BAUGH ET AL.}
\righthead{THE EPOCH OF GALAXY FORMATION}

\def\mpc {h^{-1} {\rm{Mpc}}}
\def\kpc {h^{-1} {\rm{kpc}}}
\def\msun {{\rm M}_{\odot}}
\def\ergs {{\rm erg} \, {\rm s}^{-1}}
\def\cm {{\rm cm}}
\def\kms {{\rm km} \, {\rm s}^{-1}}
\def\Hz {{\rm Hz}}

\def\arcmin {{\rm arcmin}}
\def\U {{ U_{\rm n} }}
\def\G {{ G  }}
\def\R {{\cal R  }}
\def\K {{ K  }}
\def\Omegab {{ \Omega_{\rm b}  }}

\def\lsim{\mathrel{\hbox{\rlap{\hbox{\lower4pt\hbox{$\sim$}}}\hbox{$<$}}}}
\def\gsim{\mathrel{\hbox{\rlap{\hbox{\lower4pt\hbox{$\sim$}}}\hbox{$>$}}}}

\def\and   {\rm {et al.} \rm}  
\def\etal  {\rm {et al.} \rm}

\begin{document}

\title{The epoch of galaxy formation}

\author{C.M. Baugh, S.Cole and C.S. Frenk}
\affil{Department of Physics, Science Laboratories, South Road, Durham DH1 3LE}
\author{C.G. Lacey}
\affil{Theoretical Astrophysics Center, Juliane Maries Vej 30, DK-2100 
       Copenhagen \O, Denmark}

\begin{abstract}
We use a semi-analytic model of galaxy formation in hierarchical
clustering theories to interpret recent data on galaxy formation and
evolution, focussing primarily on the recently discovered population of
Lyman-break galaxies at $z\simeq 3$. For a variety of cold dark matter
(CDM) cosmologies we construct mock galaxy catalogues subject to identical
selection criteria to those applied to the real data. We find that the
expected number of Lyman-break galaxies is very sensitive to the assumed
stellar initial mass function and to the normalization of the primordial
power spectrum. For reasonable choices of these and other model parameters,
it is possible to reproduce the observed abundance of Lyman-break galaxies
in CDM models with $\Omega_0=1$ and with $\Omega_0<1$. The characteristic
masses, circular velocities and star-formation rates of the model
Lyman-break galaxies depend somewhat on the values of the cosmological
parameters but are broadly in agreement with available data. These galaxies
generally form from  rare peaks at high redshift and, as a result, their
spatial distribution is strongly biased, with a typical bias parameter,
$b\simeq 4$, and a comoving correlation length, $r_0\simeq 4 \mpc$. The
typical sizes of these galaxies, $\sim 0.5\kpc$, are substantially smaller
than those of present day bright galaxies. In combination with data at
lower redshifts, the Lyman-break galaxies can be used to trace the cosmic
star formation history. We compare theoretical predictions for this history
with a compilation of recent data. 
The observational data match the theoretical predictions reasonably well,
both for the distribution of star formation rates at various
redshifts and for the integrated star formation rate as a function of
redshift. Most galaxies (in our models and in the data) never experience
star formation rates in excess of a few solar masses per year. Our models
predict that even at $z=5$, the integrated star formation rate is similar
to that measured locally, although less than 1\% of all the stars have
formed prior to this redshift. 
The weak dependence of the predicted star formation
histories on cosmological parameters allows us to propose a fairly general
interpretation of the significance of the Lyman-break galaxies as the first
galaxy-sized objects that experience significant amounts of star formation.
These galaxies mark the onset of the epoch of galaxy formation that
continues into the present day. The basic ingredients of a consistent
picture of galaxy formation may well be now in place.

\end{abstract}

\begin{keywords}
 {galaxies: evolution - galaxies:formation - galaxies:fundamental parameters}
\end{keywords}

\section{Introduction}

Observational studies of galaxy formation and evolution have progressed at
a breathtaking pace over the past couple of years. Data from the
refurbished Hubble Space Telescope, the Keck and other large telescopes are
now providing quantitative information on essential properties of the
galaxy population -- number densities, luminosities, colours, morphologies and
star formation rates -- over a large span of cosmic time. These data are
beginning to sketch out an empirical picture of galaxy formation and
evolution from redshift $z\simeq 4$ to the present. 

Evolution has now been established and quantified in: {\it (i)} the 
neutral hydrogen and metal content of the universe since $z\simeq 4$
(Lanzetta, Wolfe \& Turnshek 1995, Storrie-Lombardi \etal 1996, 
Wolfe \etal 1995, Lu
\etal 1996); {\it (ii)} the galaxy luminosity function since $z\simeq
1$ (Lilly \etal 1995; Ellis \etal 1996); {\it (iii)} the morphology of
field and cluster galaxies since $z\simeq 0.8$ (e.g. Abraham \etal 1996,
Dressler \etal 1994, Smail \etal 1997). The most recent addition to this remarkable list of
observational advances is the discovery of a large population of
{\it actively star-forming galaxies at $z\simeq 3$}, identified by their 
redshifted Lyman continuum breaks (Steidel \etal 1996a, 
hereafter S96). It is this
population that we are primarily concerned with in this paper.

One of the earliest windows on the physical processes at play in 
galaxy formation
is provided by studies of primordial gas clouds detected in absorption
against background quasars. The comoving density of neutral hydrogen
present in damped Lyman-alpha clouds peaks at $z\simeq 3$ when it was
comparable to the mass in baryons seen in galactic disks today
(Storrie-Lombardi \etal 1996). The decline in the abundance of neutral
hydrogen clouds seems to be accompanied by a gradual build-up of their metal
content (Lu \etal 1996). A population of bright galaxies is certainly well
established by $z=1$
(Lilly \etal 1995, Ellis \etal 1996, Kauffmann, Charlot \& White
1996). In the CFRS survey of Lilly \etal, evolution is manifest 
in a systematic variation of the shape of
the luminosity function of blue galaxies 
and a brightening of their characteristic luminosity with
increasing lookback time. The luminosity function of red galaxies, on the
other hand, seems to have changed little over this redshift interval, 
although the fraction of galaxies that have the colours of 
passively evolving ellipticals appears to fall to one third 
of its present day value by $z=1$ (Kauffmann \etal 1996).

On the whole, galaxies appear to be smaller and increasingly irregular at
higher redshift (Driver \etal 1995, Glazebrook \etal
1995a, Abraham \etal 1996, Smail \etal 1996, Odewahn \etal 1996, Pascarelle
\etal 1996, Lowenthal \etal 1997.) For example, the class of
``irregular/merger'' galaxies which are relatively rare at bright magnitudes 
 makes up about a
third to a half of all galaxies with $I_{AB} \simeq 25$. The median
redshift at this apparent magnitude (the faintest at which automated
morphological classification is possible on high resolution HST images) is
$z\simeq 0.8$. Similarly, the fraction of spirals 
in rich clusters at $z\simeq 0.5$ is higher than
in present day clusters (Dressler \etal 1994). All these studies leave
little doubt that the galaxy population has evolved significantly since 
$z=1$.

The discovery of a large population of Lyman-break galaxies at $z>3$
provides the first opportunity for statistical studies of evolutionary
processes in galaxies beyond $z=1$.  Steidel and collaborators (Steidel \&
Hamilton 1992, 1993; Steidel, Pettini \& Hamilton 1995; S96) 
searched for high redshift galaxies by selecting
objects in a colour-colour plane constructed from images in customised
$\U$, $\G$ and $\R$ band filters. For galaxies in the redshift range $3.0 <
z < 3.5$, the Lyman limit discontinuity passes through the $\U$
filter. Opacity due to intervening neutral hydrogen increases the strength of
the Lyman discontinuity regardless of the shape of the intrinsic spectral
energy distribution of the galaxy (Madau 1995). Thus, a galaxy in this
redshift range will be faint in the $\U$ band (thus becoming a
``UV-dropout'') and so will have a very red $\U-\G$ colour, whilst possibly
having a blue $\G-\R$ colour if it is undergoing significant star
formation. A similar strategy has been successfully implemented in the
Hubble Deep Field (HDF) by Steidel
\and (1996b) and Madau \and (1996). The HST $U$ filter has a shorter median
wavelength than the $\U$ filter of the ground-based observations, and so
colour selected objects in the HDF span a wider range of redshifts from $2
\lsim z \lsim 4.5$. 

Follow-up spectroscopy of the UV drop-out candidates by S96 on the
Keck telecope confirmed that these galaxies lie mostly in the expected
redshift range, $3.0 \leq z < 3.5$. Their spectra resemble those of
nearby starburst galaxies. From the apparent $\R$-band magnitude, a
dust-free model for the spectral energy distribution in the UV, and an
assumption about the initial stellar mass function (IMF), S96 inferred
star formation rates in these galaxies in the range $1-6 h^{-2} {\rm
M}_{\odot} {\rm yr}^{-1}$ for a critical density universe, where we
have expressed Hubble's constant as $H_{0} = 100 h {\rm km \, s}^{-1}
{\rm Mpc}^{-1}$.  Similarly low star formation rates have been
inferred by Lowenthal \etal (1997) for 11 galaxies in the HDF at $z=
2-4.5$. From the width of saturated interstellar absorption lines, S96
inferred tentative one-dimensional velocity dispersions in the range
$\sigma_{1D} = 180 - 320 {\rm km \, s}^{-1}$.  They concluded that the
Lyman-break galaxies they discovered could be the progenitors of the
spheroidal components of present day galaxies.

The current state of empirical knowledge on galaxy formation has been
nicely summarized by Madau \etal (1996) and Madau (1996) in the form of a
``cosmic star formation history.'' Combining a variety of surveys
(including the CFRS and the Lyman-break galaxy surveys), they derived metal
production and star formation rates as a function of redshift, from $z=0$ to
$z\simeq 5$.  Observed star formation rates over this redshift range are 
typically a few solar masses per year for individual galaxies.  
The integrated star formation rate never
differs by more than an order of magnitude over this entire redshift range,
although a peak of activity seems to have occured at $z\simeq 1$. The total
amount of metals produced by the observed populations is comparable to the
amount of metals seen in massive galaxies today, suggesting that the bulk
of the cosmic star formation has now been identified.

In this paper we employ the semi-analytic model developed in a series of
earlier papers (Cole \etal 1994; Heyl \etal 1995; Baugh \etal 1996a,
1996b), to investigate the significance of the Lyman-break galaxy
population within the context of hierarchical clustering theories of galaxy
formation. We consider the circumstances under which such a population may
form and we focus on the connection between these high redshift objects and
galaxies seen in various evolutionary stages at lower redshifts. We use the
available data to test in detail our earlier theoretical predictions for
the way in which galaxies are built up from small fluctuations in a
universe dominated by cold dark matter (CDM). Specifically, we test the
prediction of White \& Frenk (1991), Lacey \etal (1993) and 
Cole \etal (1994) that the bulk of
the stars present in galaxies today formed relatively recently, with a
median redshift of star formation of only $z \simeq 1$. The results of
these comparisons are very encouraging and suggest a general picture of
galaxy formation and evolution which is consistent with the expectations
from a broad class of hierarchical clustering cosmologies.

A more basic attempt to investigate whether the abundance of 
S96 Lyman-break galaxies is 
consistent with cold dark matter models was recently carried out by Mo \&
Fukugita (1996). They used the Press \& Schechter (1974) formalism to
calculate the number density of halos with velocity dispersion in excess
of $\sigma_{1D}= 180 \kms$, making simple assumptions about the time
required for a galaxy to form in each halo. 
They concluded that a range of 
low-density COBE-normalized CDM models are compatible with the data. 

Our semi-analytic galaxy formation scheme is briefly reviewed in
Section~\ref{s:method1}, where we discuss our procedure for generating mock
catalogues of high redshift galaxies. The colour selection criteria of S96
are reviewed in Section~\ref{s:col} and the abundance of galaxies that meet
these constraints in a variety of cosmological models is given in
Section~\ref{s:abun}.  The expected properties of high redshift galaxies --
masses, star formation rates, sizes, clustering, etc -- are presented in
Section~\ref{s:prop}.  Our models predict the entire evolutionary history
of galaxies and so, in Section~\ref{s:fate}, we illustrate the eventual
fate of a few high redshift examples and examine the statistical properties
of the descendants of the Lyman-break objects. In Section~\ref{s:epoch}, we
recast the Cole \etal (1994) predictions for the cosmic star formation
history in a manner that is directly comparable to the Madau \etal (1996)
data, and we also compare our predicted evolution in the neutral gas
fraction with observations. Finally, we present our conclusions in
Section~\ref{s:conc}.

\vspace{1cm}
\section{Semi-analytic modelling of galaxy formation}
\label{s:method1}

\subsection{General description of the model}

Semi-analytic modelling is a novel technique for calculating {\it ab initio}
the evolutionary properties of galaxies in cosmological models in which
structure forms hierarchically.  The growth of dark matter halos by
accretion and mergers is followed statistically, while physically motivated
rules are used to describe the cooling of gas in these halos, the
transformation of cold gas into stars, and the effects of feedback from
massive stars on the dynamics of the gas. The spectrophotometric properties
of the stars that form are calculated from a spectral synthesis model. The
basic physical concepts, mathematical formalisms, and first applications of
semi-analytic modelling are presented in White \& Rees (1978), Cole (1991),
Lacey \& Silk (1991), White \& Frenk (1991), Kauffmann, White
\& Guiderdoni (1993), and Cole {\it et al.}  (1994). The technique has
now been successfully applied to a variety of problems in galaxy formation
(Lacey \etal 1993; Kauffmann, Guiderdoni \& White 1994; Heyl
\etal 1995; Kauffmann 1995, 1996a,b and Baugh \etal 1996a,b, Frenk \etal 1996,
1997).

The basic rules that govern the physical processes in the galaxy formation
scheme adopted in this paper are presented in detail in Cole {\it et al.}
(1994). In brief, when a dark matter halo collapses, its associated gas is
assumed to be shock-heated and to settle into quasistatic equilibrium at
the virial temperature of the halo. This gas cools radiatively over the
lifetime of the halo and cold gas is turned into stars at a rate
proportional to the instantaneous cold gas mass. Feedback from supernovae
and stellar winds returns some of the cold gas to the hot phase, strongly
inhibiting star formation in low circular velocity halos. During a merger,
the dark matter halos coalesce, but the galaxies within them can survive
longer, eventually merging on a timescale related to the dynamical friction
time.

For this analysis, we have upgraded the Cole \etal (1994) model in various
ways.  The main modification is the replacement of the ``block model'' 
(Cole \& Kaiser 1988) as the description of the merger history of 
dark matter halos.  In the new
scheme we use a Monte Carlo method based on the analytical expression for
the halo progenitor mass function derived from the ``extended
Press-Schechter theory" (Bond \etal 1991; Bower 1991; see also Lacey \&
Cole 1993) to generate binary merger trees. Each tree describes a possible
merger history for a halo of specified final mass. At each branch in the
tree a halo splits into two progenitors, but unlike in the ``block model,''
the mass ratio of the two progenitors can take any value. This technique
enables the merger process to be followed with high time resolution, as
timesteps are not imposed on the tree but rather are controlled directly by
the frequency of mergers.  It is similar in spirit to the method used by
Kauffmann \etal (1993), but has several advantages, including that it does
not require the storage of large tables of progenitor distributions. The
new merger scheme is fully described in Lacey \& Cole (in preparation).

A further modification to the scheme is that the singular isothermal 
sphere model 
adopted by Cole \etal as a description of the dark matter halo density
profile has been replaced by the analytical form proposed by Navarro,
Frenk \& White (1996) on the basis of high resolution N-body simulations.
In the original Cole \etal model all the gas that
could cool over the entire life of the halo was assumed to be available to
form stars from the beginning of the halo lifetime. We now estimate the
supply of cold gas available to form stars by calculating the cooling rate
at a series of discrete timesteps in which successive shells of gas can
cool.

In our scheme, a cosmological model is specified by an assumption
about the nature of the dark matter together with values for the
cosmological parameters: the mean cosmic density ($\Omega_0$), the
cosmological constant ($\Lambda_0$), Hubble's constant ($H_0=100 h
{\rm km s}^{-1}{\rm Mpc}^{-1}$), 
the rms mass fluctuations in spheres of 
radius $8 h^{-1}{\rm Mpc}$ ($\sigma_{8}$), 
and the mean baryon density in units
of the critical density ($\Omegab$). 
Our galaxy formation prescription requires specifying 
6 physical parameters: (i) a star formation timescale ($\tau^{\star}_0$), (ii) a ``feedback
parameter," (iii) the shape of the initial mass function (IMF) of stars, 
(iv) an overall luminosity normalisation given by the ratio of the 
total mass in stars, including brown dwarfs, to the mass in luminous stars, 
(v) a merger timescale for galaxies, 
and (vi) the threshold mass for a galaxy merger to turn a 
disk into a spheroid (see Cole \etal 1994 and Baugh \etal 1996b 
for further details.)

The general strategy that we have adopted in this and previous papers
(Baugh \etal 1996ab, Frenk \etal 1996, 1997), is to fix the first 5 basic
parameters of the model to obtain the best possible match to the local
B-band and K-band galaxy luminosity functions and the sixth parameter to
reproduce the local relative abundances of ellipticals, S0's and
spirals. It turns out that these requirements severely restrict the allowed
range of parameter values, except for the IMF which in any of the commonly
used forms (ie. Salpeter (1955), Miller-Scalo (1979) or Scalo (1986)) has
little effect on the predicted {\it local} luminosity function.  The
evolution of the characteristic luminosity, $L_{\ast}$, however, is
sensitive to the choice of IMF, which therefore affects predictions for the
counts of faint galaxies (see Cole \etal 1994). For the most part, we have
adopted the same values of the parameters as used in the fiducial model of
Cole {\rm et al.,} allowing ourselves the freedom to use different forms
for the IMF.  We again assume that feedback is a strong function of the
halo circular velocity.  The two exceptions are that we have slightly
reduced the star formation timescale, $\tau^{\star}_0$,  
from 2 to 1.5 Gyr and we have
doubled the ratio of the galaxy merger timescale to the dynamical time in
the halo. The former change leads to an abundance of Lyman-break galaxies
in better agreement with the data (c.f. \S~3.2) while the latter change 
compensates for minor differences introduced by our new Monte-Carlo scheme
for the halo merger trees. With this choice of parameters, our new model
produces luminosity functions that are very similar to those published by
Cole \etal We have updated the original Bruzual-Charlot stellar population
synthesis model with their new version (also for solar metallicity only;
Bruzual \& Charlot 1993, Charlot, Worthey \& Bressan 1996.)

Fixing the model parameters by reference to a small subset of the data
produces a fully specified model which can then be tested against other
data, particularly high redshift data.  Thus specified, our model has
predictive power and we have presented a number of specific predictions in
earlier papers. Two of these are particularly relevant to the present
discussion. The first concerns the redshift distribution of a survey of
faint galaxies limited to magnitude $B=24$ (see figure~20 of Cole \etal
1994). Data have now been obtained by Glazebrook \etal (1995b) and by Cowie
\etal (1996). Our model predictions are in good agreement with these
data, as may be seen in Fig.~1 of Frenk \etal (1997) 
(see also Fig.~15 of White \& Frenk 1991 and Kauffmann, Guiderdoni \& 
White 1994). 
The second prediction
to which we will return in Section~5 of this paper, is the 
cosmic star formation history, presented in Figure~21 of Cole \etal (1994) and
in Fig. 14 below.

\begin{figure*}
{\epsfxsize=18.truecm \epsfysize=21.truecm 
\epsfbox[0 100 580 720]{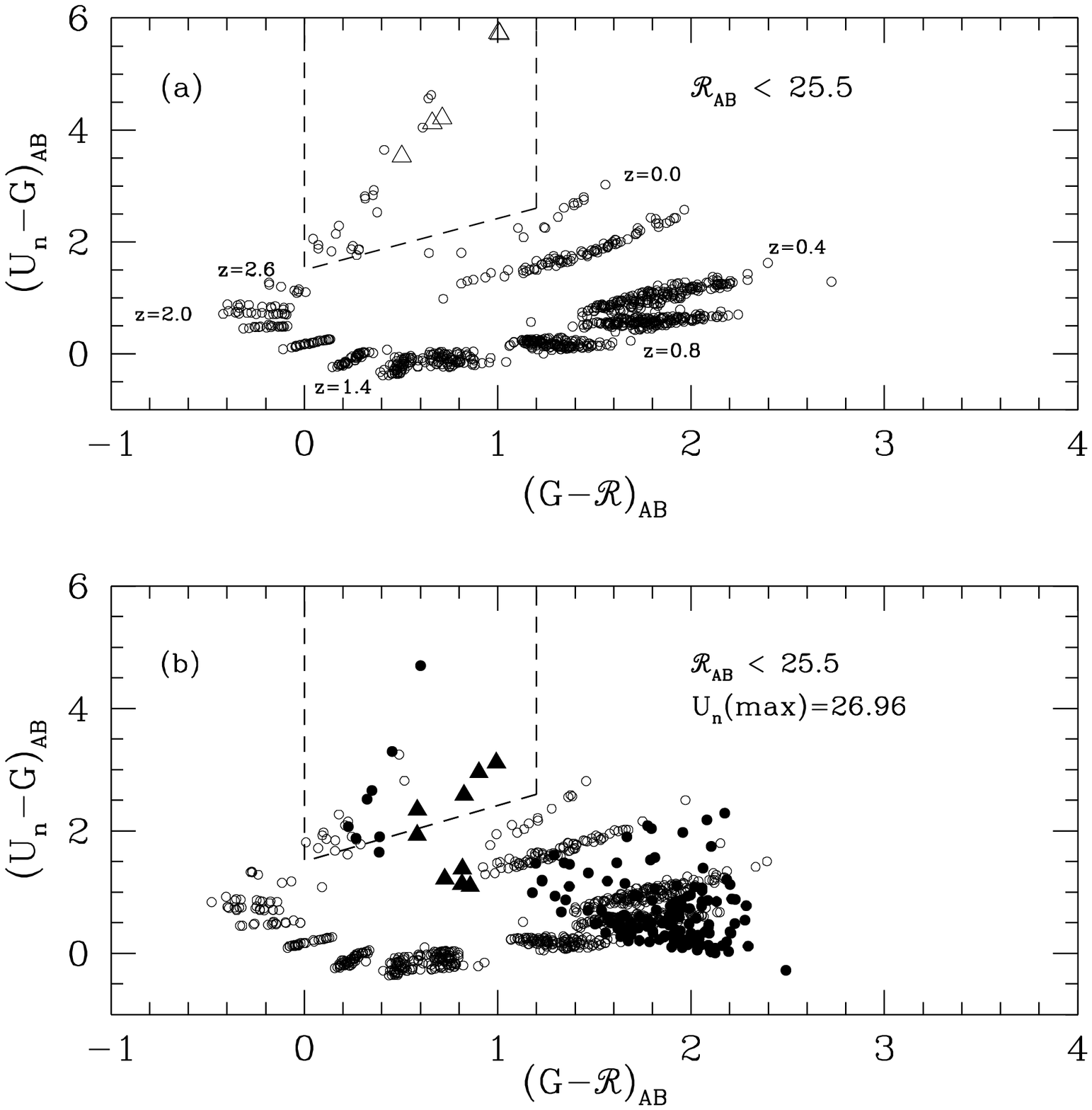}}
{\small {\bf FIG. 1}
Colour-colour diagrams for galaxies in the standard CDM model. Galaxies
brighter than $\R_{\rm AB}=25.5$ are plotted; the area of the mock field sampled is
$14.6$ square arcminutes.
(a) $\U-\G$ colours computed from the true $\U$ magnitude for all
galaxies. The open circles show galaxies with redshifts $z< 3.0$ and the open
triangles galaxies with redshifts in the range $3.0 < z < 3.5$. Model
results are output at specific redshifts and this is reflected in the
discrete regions populated in the diagram. The redshifts of selected
outputs are indicated. 
(b) $\U-\G$ colours computed by setting the $\U$ magnitude to a
detection limit of $\U = 26.96$ whenever $\U$ is fainter than the detection
limit. Galaxies in this class are denoted by filled symbols. As in panel
(a), the triangles denote galaxies with redshifts in the range $3.0 < z < 3.5$.
}
\label{fig:ug_gr}
\end{figure*}

\subsection{Modelling galaxies at high redshift}

We begin by constructing merger trees for a grid of halo masses,
specified at some redshift $z_{\rm halo}$. Typically, we generate between 
5 and 20 different realizations for each mass, depending upon the 
Press-Schechter abundance.  The galaxy formation rules are
applied along the branches of each halo tree, starting at the highest
redshift of interest and propagating through to $z_{\rm
halo}$. Volume-limited samples or redshift catalogues are generated
from the model output by weighting the galaxies in a halo tree of a given
mass by its predicted Press-Schechter abundance at $z_{halo}$. Mock
catalogues consisting of galaxies selected according to any colour-magnitude 
criteria can be readily generated from the model output.

To construct mock catalogues with the selection criteria of S96, we
calculate broad band colours for the model stellar populations using
the same set of customised filters employed by S96. (The filter functions were
kindly provided by C. Steidel.) The dominant effect that determines
whether a galaxy is a UV dropout is absorption of the galactic UV light by
intervening cold gas. We calculate the effect of absorption on the
spectral energy distributions (SEDs) of high redshift galaxies using the
procedure developed by Madau (1995). In this way, we are able to select
model galaxies according to exactly the same colour criteria as applied to
the observational data by S96.

For most of this study we have generated trees starting at $z_{\rm
halo}=2.6$. A high starting redshift is desirable in order to minimize
inaccuracies at the high mass end of the mass distribution of
progenitors introduced when the Monte-Carlo scheme is applied over a
large range of expansion factor (Lacey \& Cole, in preparation). We
have checked, however, that our results are insensitive to the exact
choice of $z_{\rm halo}$. To obtain the expected total number of
galaxies in the apparent magnitude range observed by Steidel et al.,
$\R_{\rm AB} < 25.0$, we also generated mock galaxy samples from a
grid of halo masses laid down at $z_{\rm halo}=0$. At $\R_{\rm AB} < 25.0$, the
median redshift is $z \sim 0.7$.

\section{ Model results}

In this section we investigate the properties of the Lyman-break galaxies
that form in our models and, wherever possible, we compare these to the
properties of the real Lyman-break galaxies discovered by
S96. Specifically, we consider the number density, masses, star formation
rates and sizes of these galaxies and we present predictions for their
clustering properties. To fully specify a model we need to adopt values for
the cosmological parameters, $\Omega_0$, $\Lambda_0$, $h$ and $\sigma_8$,
and also a value for the baryon fraction, $\Omegab$, and an IMF. There are
considerable uncertainties in these choices. We carry out calculations in
three different cosmological models: the standard CDM model ($\Omega_0=1$,
$h=0.5$, $\sigma_8=0.67$); a flat, low-density CDM model ($\Omega_0=0.3$,
$\Lambda_0=0.7$, $h=0.6$, $\sigma_8=0.97$) and an open CDM model
($\Omega_0=0.4$, $h=0.6$, $\sigma_8=0.68$). These parameters are typical of
those favoured by large-scale structure constraints. For example, all of our
models produce approximately the correct abundance of galaxy clusters at
the present day and, under standard assumptions, the low-density models
also match the 4-year COBE microwave background anisotropy data (Bennet
\and 1996; Liddle \etal 1996; White, Efstathiou \& Frenk 1993; Eke, Cole \&
Frenk 1996; Viana \& Liddle 1996; Cole et al. 1997.)  For our standard
$\Omega_0=1$ cosmology, we vary the normalisation of the primordial
fluctuation spectrum, considering both the above value, $\sigma_8=0.67$,
adopted in the fiducial model of Cole \etal (1994), and the lower value,
$\sigma_8=0.5$, preferred by Eke, Cole \& Frenk (1996) for consistency with
the observed cluster X-ray temperature function.

We consider models with two different baryon fractions, $\Omegab h^2=0.015$
and $\Omegab h^2=0.030$. The first agrees with the estimate by Copi, Schramm 
\& Turner (1996) from Big Bang nucleosynthesis and 
the second is consistent with
the claim of Tytler, Fan \& Burles  (1996) of a low primordial deuterium abundance in gas clouds
at high redshift. We consider three possibilities for the IMF, all of which
are consistent with solar neighbourhood data, given the uncertainties in
its past star formation history: Miller-Scalo (1979), Scalo (1986) and
Salpeter (1955) (see figure~4 in Cole \etal 1994 for the specific
parametrizations used).  The parameters of these models (and of variants
considered below) are summarized in Table~1.

\subsection{Two-colour selection}
\label{s:col}

We first consider the broad-band colours of our model galaxies and test the
assumption that high redshift galaxies can be efficiently identified from
their location in the $\U-\G$ {\it vs} $\G-\R$ colour-colour plot
constructed by S96. The analysis of Steidel, Hamilton \& Pettini (1995)
suggests that galaxies with redshifts in the range $3.0 < z < 3.5$ should
lie within the trapezium bounded by the dashed lines in
Fig. 1.  At these redshifts, the Lyman break passes through
the observer's frame $\U$ band.

Our predicted colour-colour diagram, for the case of standard CDM, is
shown in Fig. 1a. The data plotted correspond to an area
of $14.6$ square arcminutes, equal to the area of the Q0347-3819 field
observed by Steidel, Pettini \& Hamilton (1995).  (For clarity the
points have been given small random displacements in the $x$ and $y$
directions. The localizations of the points in bands reflect the
discrete set of output redshifts.) The open triangles in
Fig. 1a indicate galaxies that have redshifts in the
range $3.0 < z < 3.5$.  As anticipated by Steidel \etal (1995), all
galaxies with $3.0 < z < 3.5$ lie inside the trapezium in
Fig. 1a. However, our models show some contamination by
galaxies with redshifts in the range $2.7 < z < 3.0$ that congregate
near the bottom-left corner of the trapezium.  This contamination is,
in fact, consistent with the evolutionary tracks shown in figure~2 of
Steidel \and (1995).

Arbitrarily deep Lyman breaks cannot be measured in practice because
the observations are subject to a detection limit in the $\U$
band. For the Q0347-3819 field, the $3 \sigma$ detection limit is
$\U=26.96$. 
Fig. 1b illustrates the effect of imposing
such a detection limit on the appearance of the colour-colour
plot. Whenever a galaxy has a true $\U$ magnitude fainter than the
field limit, we plot it at the assumed $\U$ limit using a filled
symbol. A significant fraction of high redshift galaxies now lie below
the trapezium region, confirming the remark by Steidel \and (1995)
that their selection of candidates is likely to be an underestimate of
the true abundance.

Comparison of Fig. 1b with figures~4, 7, 10 and 13 of Steidel
\etal (1995) shows that our models tend to produce too few galaxies with
$(\U-G)$ and $(G-\R)$ less than unity. Galaxies in this part of the
observational diagram are likely to be predominantly faint foreground dwarf
irregulars. The discrepancy may be due in part to our use of 
solar metallicity stellar populations.
This is also apparent in the comparison between our
models and the local luminosity function determined by Lilly \and (1995)
from the CFRS survey (See figure 16 of Baugh, Cole \& Frenk 1996b.) Beyond
$z \gsim 0.2$, however, the luminosity functions of our models for both red
and blue galaxies agree reasonably well with the CFRS data.

\centerline{{\epsfxsize=8.6truecm \epsfysize=8.6truecm 
\epsfbox[30 150 580 720]{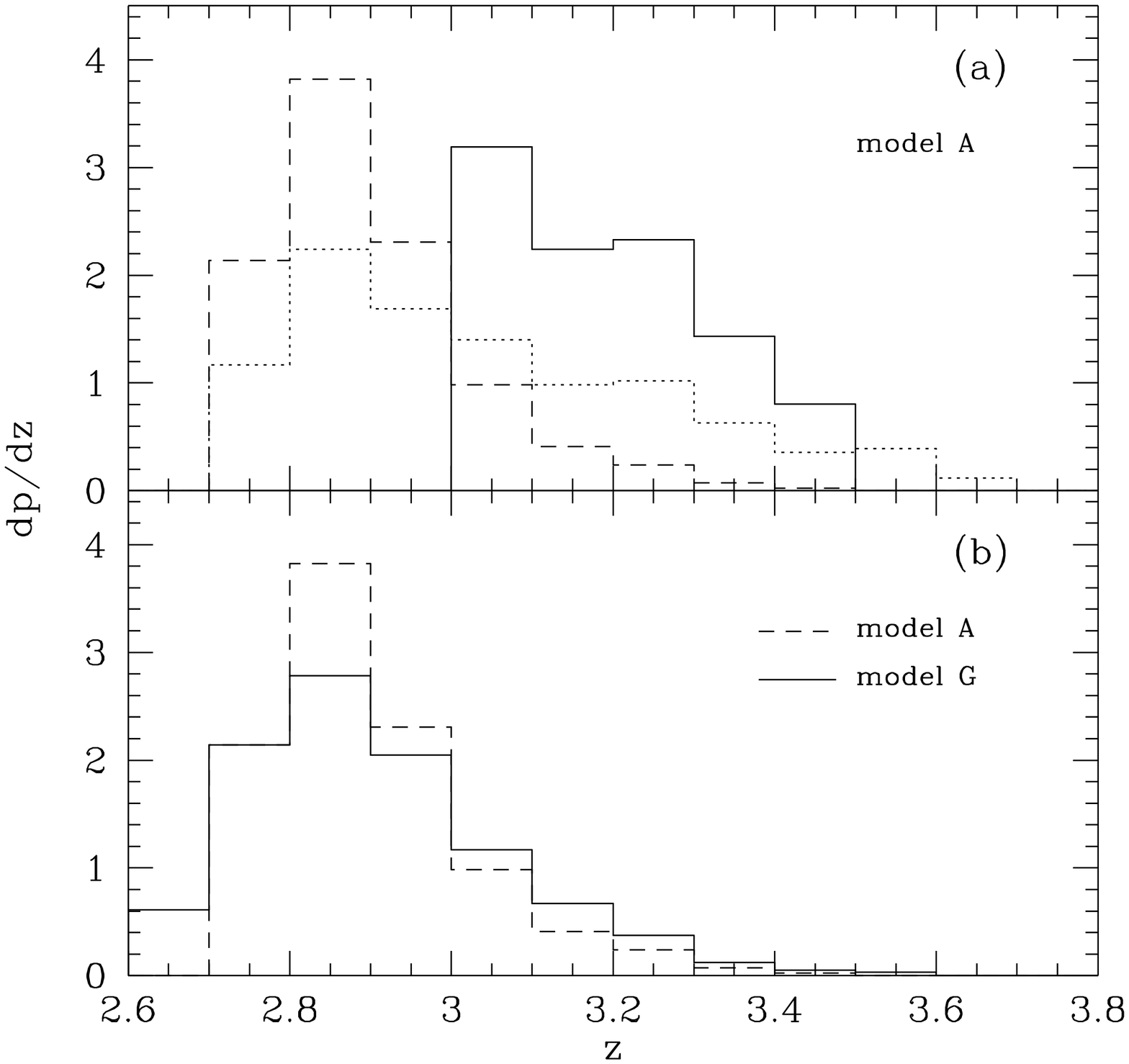}}}
{\small {\bf FIG. 2} 
Predicted redshift distributions for galaxies brighter than apparent
magnitude $\R_{\rm AB}=25.0$ selected in various ways. 
Panel (a) shows results for our standard CDM model A. The
solid line shows the distribution for galaxies with redshift in the
range $3.0 < z < 3.5$. The dotted line shows the distribution for
galaxies that meet the colour criteria of Steidel \and (1995). The
dashed line refers to galaxies that meet the Steidel \and (1995)
colour selection, after the detection limit in $\U$ for a typical
observed field has been applied - model galaxies fainter than this in $\U$ are 
assigned this limiting magnitude. In panel (b) the dashed histogram is
repeated from panel (a) and the solid histogram shows the
corresponding redshift distribution of galaxies brighter than $\R_{\rm
AB}=25$ in the low-$\Omega$ model~G brighter than 
$\R_{\rm AB}=25$ that satisfy the Steidel \etal
colour selection criteria, taking into account the typical field limit in $\U$.
These and subsequent histograms are normalized so that they enclose
unit area.  
\label{fig:dndz}
}

\subsection{The Abundance of high redshift galaxies}
\label{s:abun}

\begin{figure*}
{\epsfxsize=16.truecm \epsfysize=16.truecm 
\epsfbox[-100 150 580 720]{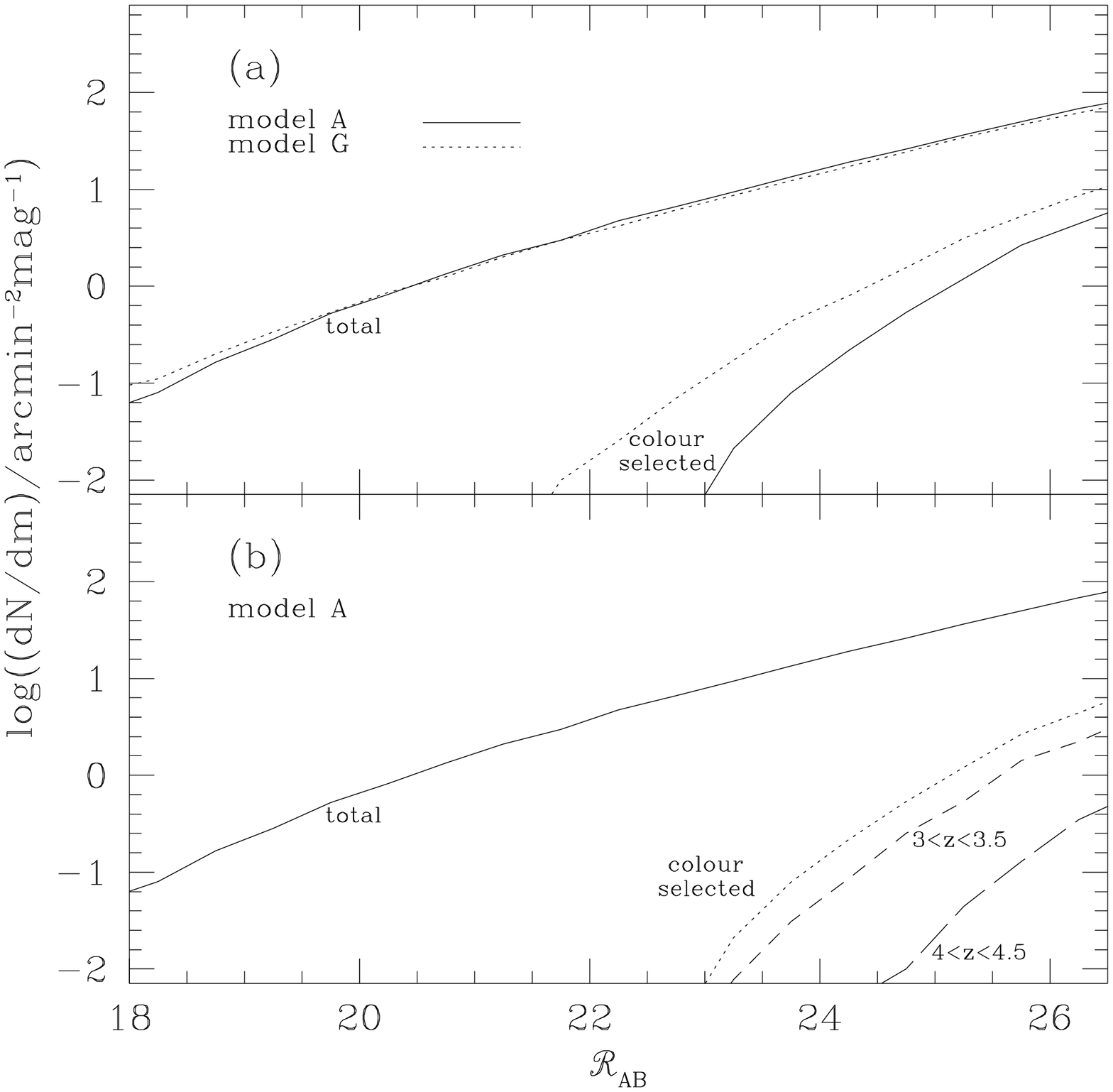}}
{\small {\bf FIG. 3}
Predicted galaxy number counts. In panel (a), the higher amplitude pair of
lines shows the total counts in model A (solid line) and model~G (dotted
line). The lower amplitude pair of lines shows the counts of galaxies in
each of these models that satisfy the Steidel \etal colour selection
criteria alone, without any constraint on their $\U$ magnitudes.
In panel (b) the total counts in model A are again shown by the
solid line. The dotted line now shows the counts of colour selected
galaxies; the short dashed line shows the counts of galaxies with
redshifts in the range $3.0 < z < 3.5$; and the long-dashed line shows the
counts of galaxies with redshifts in the range $4.0 < z < 4.5$.
\label{fig:counts}
}
\end{figure*}

\begin{table*}
\begin{center}
\caption[dummy]{Model parameters. Columns 2-6 give the values of the
cosmological parameters defined in the text. Column~7 gives, $\Upsilon$, 
the ratio of
the total mass in stars to the mass in luminous stars. Column~8 indicates the
stellar initial mass function (IMF) used.}
\vspace{1cm}
\begin{tabular}{cccccccl}
\hline
\multicolumn{1}{l} {model} & 
\multicolumn{1}{l} {$\Omega_0$} & 
\multicolumn{1}{l} {$\Lambda_0$} & 
\multicolumn{1}{l} {$h$} & 
\multicolumn{1}{l} {$\sigma_{8}$} & 
\multicolumn{1}{l} {$\Omegab$} & 
\multicolumn{1}{l} {$\Upsilon$} & 
\multicolumn{1}{l} {IMF} \\
\hline
 A    &   1.0    &    0.0     & 0.5 &  0.67 & 0.06  & 2.8   & Miller-Scalo \\
 B    &   1.0    &    0.0     & 0.5 &  0.67 & 0.06  & 2.3   & Scalo \\
 C    &   1.0    &    0.0     & 0.5 &  0.67 & 0.06  & 1.5   & Salpeter \\
 D    &   1.0    &    0.0     & 0.5 &  0.50 & 0.06  & 2.5   & Miller-Scalo \\
 E    &   1.0    &    0.0     & 0.5 &  0.67 & 0.12  & 6.4   & Miller-Scalo \\
 F    &   0.4    &    0.0     & 0.6 &  0.68 & 0.04  & 2.8   & Miller-Scalo \\
 G    &   0.3    &    0.7     & 0.6 &  0.97 & 0.04  & 2.4   & Miller-Scalo  \\
 H    &   0.3    &    0.7     & 0.6 &  0.97 & 0.08  & 4.2   & Miller-Scalo \\
\hline
\label{tab:models}
\end{tabular}
\end{center}
\end{table*}

\begin{table*}
\begin{center}
\caption[dummy]{The abundance of high redshift galaxies per square degree
brighter than $\R_{\rm AB} = 25$ for the models listed in 
Table 1. 
The quoted numbers are derived from 10 realisations of a 
catalogue each covering $0.1$ square degrees.
The final row gives results from the observational study of S96 }
\vspace{1cm}
\begin{tabular}{cccccccc}
\hline
\multicolumn{1}{c} {model} & 
\multicolumn{1}{c} {${\cal N }(R_{AB} < 25.0)$} & 
\multicolumn{1}{c} {${\cal N}(3.0 < z < 3.5)$} & 
\multicolumn{1}{c} {${\cal N}({\rm colour})$}  & 
\multicolumn{1}{c} {${\cal N}({\rm colour}+ \U)$} & 
\multicolumn{1}{c} {\% of ${\cal N} (\R_{\rm AB} < 25.0)$} \\ 
\hline 
 A  &  153 000  & \hphantom{0}890 & 2000 & 1000  & \hphantom{0}0.7 \\ 
 B  &  108 000  &  \hphantom{00}50 &  \hphantom{0}130 &   \hphantom{00}60  & \hphantom{00}0.05\\
 C  &  142 000  &   1100 & 2400 & 1300  & \hphantom{0}0.9\\
 D  &  117 000  &   \hphantom{00}60 &  \hphantom{0}180 &  \hphantom{0}110  & \hphantom{00}0.09\\
 E  &  152 000  &  1100 & 2100  & 1100  & \hphantom{0}0.7\\
 F  &  101 000  & 1100 & 2200 & 1000 & \hphantom{0}1.0\\
 G  &  126 000  & 2400 & 5100 & 2900  & \hphantom{0}2.3\\
 H  &  158 000  & 4200 & 8500 & 4700  & \hphantom{0}3.0\\
 observed &  110 000 & --  & --  &   $1400 \pm 300$ & 
 $1.3\pm 0.3$\hphantom{0}    \\
\hline
\label{tab:prop}
\end{tabular}
\end{center}
\end{table*}

Steidel, Hamilton \& Pettini (1995) defined a ``robust candidate'' for a
Lyman-break galaxy in the redshift range $3.0 < z < 3.5$ to be an object
brighter than $\R_{\rm AB} = 25$, with $\U-\G$ and $\G-\R$ colours in the
trapezium region of Fig.~1 and which is undetected in the $\U$ band. They
estimated the surface density of robust candidates to be $0.40 \pm 0.07
\arcmin^{-2}$, corresponding to $1.3 \%$ of their total counts brighter
than $\R_{\rm AB} = 25$. These counts are $30~\arcmin^{-2}$, with
a Poisson uncertainty of 2\%. Brainerd \etal (1995) quote a value of
$47~\arcmin^{-2} $ for the counts to the same magnitude. The
difference seems to stem from different incompleteness corrections and
uncertainties in the conversion from aperture to total magnitudes.

We now examine which, if any, of our hierarchical clustering models
produces an acceptable abundance of Lyman-break galaxies.  The parameters
of the models we have investigated are summarized in columns~2-6 of
Table 1. The seventh column gives the luminosity
normalization of each model, $\Upsilon$, which is defined as the ratio of
the total mass in stars formed in the model to the mass formed in luminous
stars {\it i.e.} stars with mass greater than $0.1M_{\odot}$. 
This parameter is set in all cases to match the knee of the observed
local field galaxy luminosity function, as described by Cole \and (1994).

Table 2 gives the abundance of galaxies predicted by our
models. Where available the observed values are shown in the bottom row
of the table. The second column, ${\cal N }(R_{AB} < 25.0)$, gives the
total number of galaxies brighter than $\R_{\rm AB} = 25$. The next
three columns demonstrate the effect of applying various selection criteria
to this $\R_{\rm AB} < 25$ sample. The third column, ${\cal N}(3.0 < z
< 3.5)$, gives the number of galaxies per square degree with redshift
in the range $3.0 < z < 3.5$.  The fourth column, ${\cal N}({\rm
colour})$, gives the number of galaxies in the region of the
colour-colour diagram from which Steidel \and (1995) selected their high
redshift candidates. As shown in Fig. 1, a significant
fraction of these galaxies are at redshifts just below 3.  The fifth
column, ${\cal N}({\rm colour}+ \U)$, is the number of galaxies
remaining in the colour selected region after the $\U\leq26.96$
magnitude limit for the Q0347-3819 field is applied and model galaxies 
fainter than this have had a $3 \sigma$ lower limit assigned for 
their $\U - G$ colour. This removes
roughly half of the high redshift candidates in column~4 from the
colour selected region. The final column gives the fraction of
galaxies that meet the colour selection criteria, after applying a
typical field limit in $\U$, as a percentage of the total counts
brighter than $\R_{\rm AB}=25.0$
 
It is clear from Table 2 that the abundance of high redshift
galaxies expected in a given cosmology is very sensitive to the adopted IMF
and to $\sigma_8$, the normalisation of the primordial power spectrum. For
example, replacing the Scalo by the Miller-Scalo IMF in our fiducial
standard CDM model, produces an increase of nearly a factor of 20 in the
number of high redshift galaxies listed in column~5. This sensitivity
arises because, when normalized to the same total mass in luminous stars,
the Miller-Scalo IMF contains several times more stars of around 
$10 M_{\odot}$ than the Scalo IMF. 
Similarly, increasing $\sigma_8$ in the $\Omega_0=1$
model from 0.5 to 0.67 increases the number of high redshift galaxies by a
factor of $10$. This dependency reflects the fact that the brightest
Lyman-break galaxies at $z\simeq 3$ tend to come from the tail of rare
objects in the mass distribution at this redshift. With our procedure for
normalizing the luminosity of the models, the predicted abundances are
insensitive to the value of $\Omegab$. However, the precise value of
$\Upsilon$ does affect the abundances; reducing $\Upsilon$ (at the expense
of a poorer match to the local luminosity function) would boost the number
of high redshift galaxies.

\begin{table*}
\begin{center}
\caption[dummy]{
Comparison of the abundance of high redshift galaxies in models 
A and G with the abundances inferred from the Hubble Deep Field 
by Madau \etal (1996).
Madau \etal estimate that the $U_{300}$-band dropouts lie in the 
redshift range $2.0 < z < 3.5$ and the $B_{450}$-band dropouts in
the redshift range $3.5 < z < 4.5$. 
This Table refers to $U_{300}$-band 
dropouts, whilst Table 4 refers to $B_{450}$-band dropouts. 
The values in the table give the number of objects per square degree.
The final column gives the number of colour selected objects as 
a percentage of the total number of objects in the sample.
}
\vspace{1cm}
\begin{tabular}{cccccc}
\hline
\multicolumn{1}{c} {model} & 
\multicolumn{1}{c} {${\cal N}_{tot}(B_{450}<26.8)$} & 
\multicolumn{1}{c} {${\cal N}(2.0 < z < 3.5)$} & 
\multicolumn{1}{c} {${\cal N}({\rm colour})$}  & 
\multicolumn{1}{c} {\% of ${\cal N}_{tot}$} \\ 
\hline 
 A       &  $360 \times 10^3$  & $54 \times 10^3$ & $66 \times 10^3$ & 18 \\
 G       &  $290 \times 10^3$  & $65 \times 10^3$ & $79 \times 10^3$ & 27 \\
observed &  $(320 \pm 16) \times 10^3  $ & - & $(46 \pm 6) \times 10^3 $ & 14\\
\hline
\label{tab:hst}
\end{tabular}
\end{center}
\end{table*}

\begin{table*}
\begin{center}
\caption[dummy]{
Comparison of the abundance of high redshift galaxies in models 
A and G with the abundances inferred from the Hubble Deep Field 
by Madau \etal (1996) for $B_{450}$-band dropouts.
}
\vspace{1cm}
\begin{tabular}{cccccc}
\hline
\multicolumn{1}{c} {model} & 
\multicolumn{1}{c} {${\cal N}_{tot}(V_{606}<27.7)$} & 
\multicolumn{1}{c} {${\cal N}(3.5 < z < 4.5)$} & 
\multicolumn{1}{c} {${\cal N}({\rm colour})$}  & 
\multicolumn{1}{c} {\% of ${\cal N}_{tot}$} \\ 
\hline 
 A       &  $790\times 10^3$ & $9.9\times 10^3$  & $5.9\times 10^3$ & 0.7 \\
 G       &  $620\times 10^3$   & $16\times 10^3$  & $9.4 \times 10^3$ & 1.5 \\
observed &  $(620 \pm 20) \times 10^3$ & - & $(9.3 \pm 2.5) \times
 10^3$ & 1.5 \\  
\hline
\label{tab:hst2}
\end{tabular}
\end{center}
\end{table*}

The conclusion to be drawn from Table 2 is that it is possible
to reproduce the observed number of Lyman-break galaxies, $1400
\pm 300$ per square degree, in a variety of cosmological models by reasonable
adjustments to the input parameters. For example, within the observational
errors, the standard CDM model produces the required abundance if the
Miller-Scalo (Model~A) or the Salpeter (Model~C) IMF is assumed. Increasing
the characteristic star formation timescale, $\tau^{\star}_0$, from 1.5 Gyr
to 2 Gyr decreases the number of Lyman-break galaxies by about a factor
two.  Similarly, the open model F with a Miller-Scalo IMF is
acceptable. The flat low-$\Omega$ models~G and~H produce about 2 to 3 times
as many Lyman-break galaxies as observed.  These abundances would be
reduced, however, if the Miller-Scalo IMF were replaced by the Scalo IMF or
if $\tau^{\star}_0$ were increased. Model D, the standard CDM cosmology
with density fluctuations normalised to reproduce the observed abundance of
rich clusters (Eke \etal 1996) produces far too few Lyman-break
galaxies. However it would be premature to conclude that a standard CDM
model with this normalization is incompatible with the high redshift data.
For example a simple modification of the Cole \etal (1994) star formation
rules, in which the timescale $\tau^{\star}_0$ is scaled with the dynamical
time of the galaxy ($\tau^{\star}_0\propto (1+z)^{3/2}$) instead of
remaining constant with redshift, results in an increase of ${\cal N}({\rm
colour}+ \U)$ from $80$ to $1900$, without significant change in the
properties of galaxies at the present time. We plan to explore the effects
of such variations in the modelling of star formation and feedback in a
future paper (Cole \etal in preparation).  Note that not all the models
listed in Table 2 reproduce the total counts of galaxies
brighter than $\R_{\rm AB}=25$ quoted by Steidel \and (1995). However, as
mentioned above, these counts may be uncertain by a significant factor.

Our predicted redshift distribution for Lyman-break galaxies after the
various selection criteria have been applied is shown in Fig. 2. 
Only galaxies brighter than  $\R_{\rm AB} = 25$ are included in
this plot. As was evident from Fig.~1, the Steidel \etal (1995) colour
criteria allow a significant population of galaxies with redshifts just
below 3. Introducing a $\U$-band detection limit 
biases the sample against the highest redshift galaxies whose light 
experiences the most absorption by intervening cold gas, skewing the 
distribution of robust candidates towards $z \sim 3$. The
top panel of Fig. 2 shows results for our standard CDM
model A and the bottom panel compares these with results from the flat
low-$\Omega$ model~G. The two distributions, heavily constrained by the
selection criteria, are very similar.

Fig. 3 shows the number counts of galaxies predicted in
two of our models. In Fig. 3a, the solid lines refer to
model A and the dotted lines to model~G. The higher amplitude pair of
curves gives the total number counts in these two models while the lower
amplitude curves give the number of objects that satisfy the Steidel \and
$\U-\G$, $\G-\R$ colour criteria. The fraction of the total counts represented
by Lyman-break galaxies increases rapidly with increasing magnitude. The
counts in model A are shown in more detail in Fig. 3b.
Again, the high amplitude solid curve gives the total counts while the
dotted line shows the number of galaxies that satisfy the colour selection
criteria. The counts of galaxies with redshifts in the range
$3.0 < z < 3.5$ are indicated by the short-dashed line and the counts at 
$4.0 < z < 4.5$ are shown by the long dashed line. The latter are
lower by about one order of magnitude. 

Steidel \etal (1996b) and Madau \etal (1996) have applied a similar
technique to isolate high redshift galaxies in the Hubble Deep Field
(Williams \etal 1996).  The HDF was imaged in four passbands and so
two-colour selection can be applied to select galaxies that `drop out' in
two passbands, $U_{300}$ and $B_{450}$.  The $U_{300}$ dropouts are
predicted to lie in the redshift range $2.0 < z < 3.5$ whilst the $B_{450}$
dropouts should have redshifts between $3.5 < z < 4.5$ (Madau \etal 1996).
We have made mock HDF catalogues from the output of our model, using the
same filters and applying the detailed colour selection given by Madau
\etal A comparison of the abundance of high redshift objects with the
inferred abundances for the HDF is given in Tables 3 and
4.  In Table 3 we consider galaxies brighter than
$V_{606} = 28.0$ and $B_{450}=26.8$, while, in Table 4, we
consider galaxies brighter than $V_{606} = 27.7$.  These data lead to a
similar conclusion as the data in Table~2: several of our models 
(as did the some of the more successful models of White \& Frenk 1991) 
predict approximately the observed abundance of high redshift galaxies 
thoughout the redshift range $2.0 < z < 4.5$. 
Our predicted abundances
are sensitive to the IMF assumed -- model B with a Scalo IMF gives seven
times fewer $B_{450}$ dropouts compared with model A which uses a
Miller-Scalo IMF.

\subsection{Properties of high redshift galaxies}
\label{s:prop}

In this Section we consider the properties of galaxies in models 
A and G that satisfy the Steidel \etal colour selection criteria 
and that are brighter than $\R_{\rm AB} = 25.0$; we do not 
apply any conditions on their $\U$ magnitudes.

\subsubsection{Dark matter halos}

The masses of the dark matter halos that 
harbour Lyman-break galaxies brighter than $\R_{\rm AB}=25.0$ 
and the circular velocities of the halos in which these galaxies formed 
are plotted in Fig. 4. 
The solid lines correspond to the standard CDM model A, and the dashed
lines to the flat, low-$\Omega$ model~G. The halo masses plotted in
the top panel are remarkably similar in the two cosmologies.  This is
largely a coincidence arising from the interplay between the selection
criteria imposed on the galaxies and the overall halo mass
distributions in the two cosmologies. The circular velocities plotted
in the bottom panel are also similar, with a shift towards lower
values in the low-$\Omega$ cosmology.

S96 estimated velocity dispersions for the Lyman-break galaxies from the
widths of heavily saturated interstellar absorption lines.
As they point out, these measurements may be contaminated by turbulent motions
in the gas. Alternatively, they may be due entirely to gravitationally
supported random motions and, in this case, their measurements indicate
velocity dispersions in the range $\sigma_{1D} = 180 - 320 \, {\rm km \,
s}^{-1}$, corresponding to circular velocities
$V_c=\sqrt{2}\sigma_{1D}=250-450 \, {\rm km \, s}^{-1}$. If the line widths
are due to rotational motion in a disk of constant circular velocity,
$V_c$, then the observed range of full-width at half-maximum, $400 - 700\,
\kms$, corresponds to $V_c \approx 250 - 430 \,
\kms$ for randomly oriented disks. Our model predictions in both 
cosmologies, illustrated in Fig. 4, are consistent with
these numbers. Note, however, that the circular velocities plotted in the
Figure are asymptotic halo values. The actual circular velocity is a
function of radius.
This, as well as the redistribution of mass associated with 
the formation of the galaxy, will affect what can be measured 
observationally.
In principle, these can be substantially different from the 
asymptotic halo values. We intend to explore this 
issue in detail in a subsequent paper.

\centerline{
{\epsfxsize=8.8truecm \epsfysize=8.5truecm 
\epsfbox[0 10 580 750]{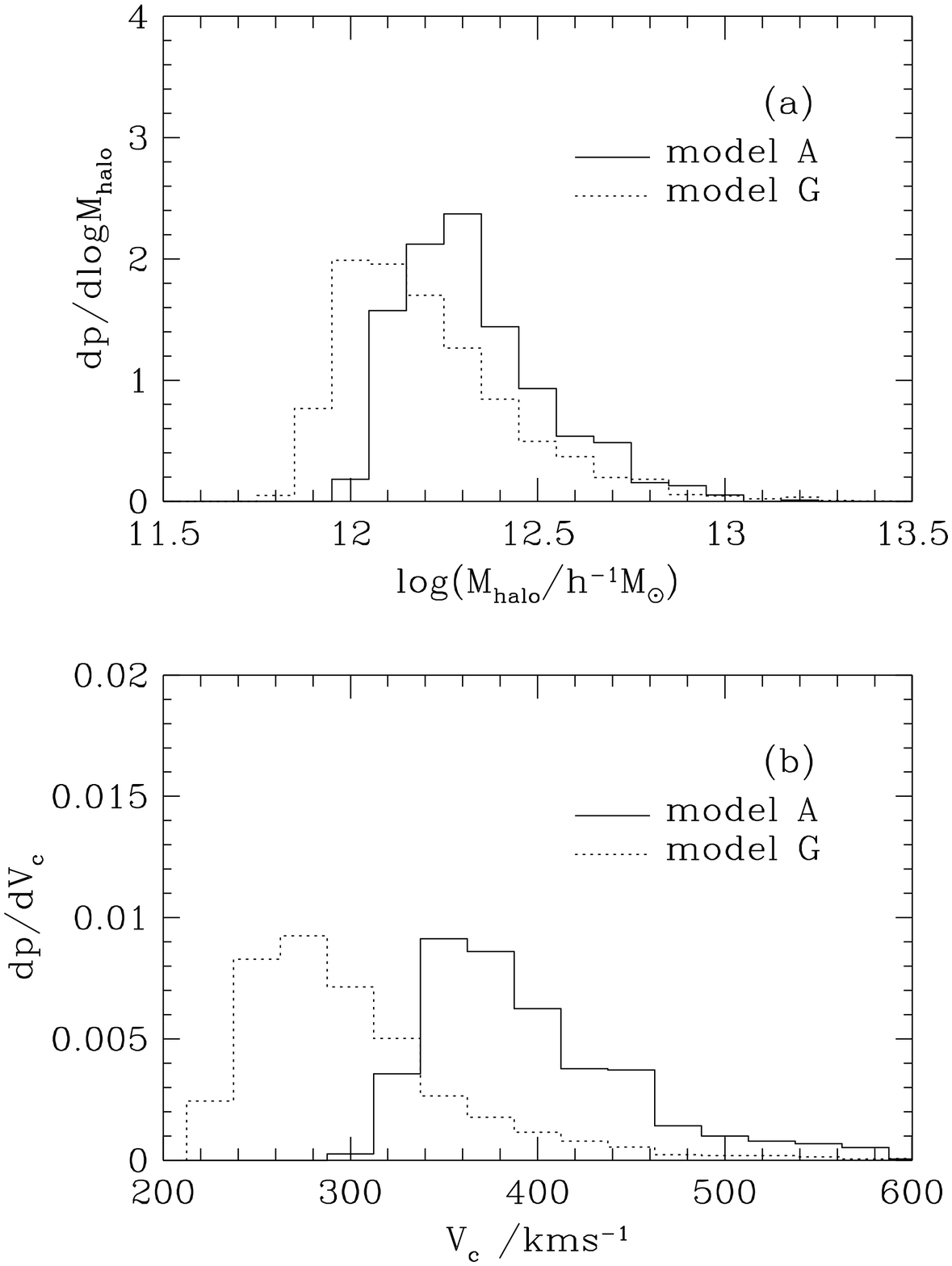}}
}
{\small {\bf FIG. 4} 
Masses and circular velocities of the dark matter halos that harbour
Lyman-break galaxies. Galaxies brighter than $\R_{\rm AB} = 25.0$
satisfying the colour criteria of Steidel \etal (1995) are
included. The solid lines show the
distributions for the standard CDM model A, whilst the dotted lines
show distributions for the flat low-$\Omega$ model~G. The top panel
gives the distribution of halo masses, and the bottom panel the
distribution of circular velocities for the halos in which each galaxy formed.  
\label{fig:dmhalo}
}

\subsubsection{Stellar masses and star formation rates}

The stellar masses of our model Lyman-break galaxies brighter than
$\R_{\rm AB}=25.0$ are plotted in Fig. 5. As before,
the solid line shows results for the $\Omega_0=1$ model
A and the dotted line for the flat, low-$\Omega$ model~G. The
stellar masses are typically three times larger in the low-$\Omega$
cosmology. This difference is the result of the selection criteria
imposed on these galaxies: luminosity distances are larger in the low
density model, so galaxies selected at a given apparent magnitude
limit must have larger luminosities, and thus larger stellar masses.

An indication of the stellar masses of real Lyman-break galaxies comes from
$K$-band imaging of 5 candidate UV dropouts carried out by S96 at the
Keck telescope. For these 5 candidates, they find $K_{\rm AB} = 23.2 -
24.0$, and colours $0.4 \leq (\R_{\rm AB}-K_{\rm AB}) \leq
1.3$. Galaxies in our models have $K$ magnitudes in the range $K_{\rm
AB} = 22 - 24$, with colours in the range $0.5 \leq \R_{\rm AB}-K_{\rm
AB} \leq 1.4$, in excellent agreement with the data, as shown in 
Fig. 6.

\centerline{
{\epsfxsize=8.8truecm \epsfysize=9.truecm 
\epsfbox[50 400 580 720]{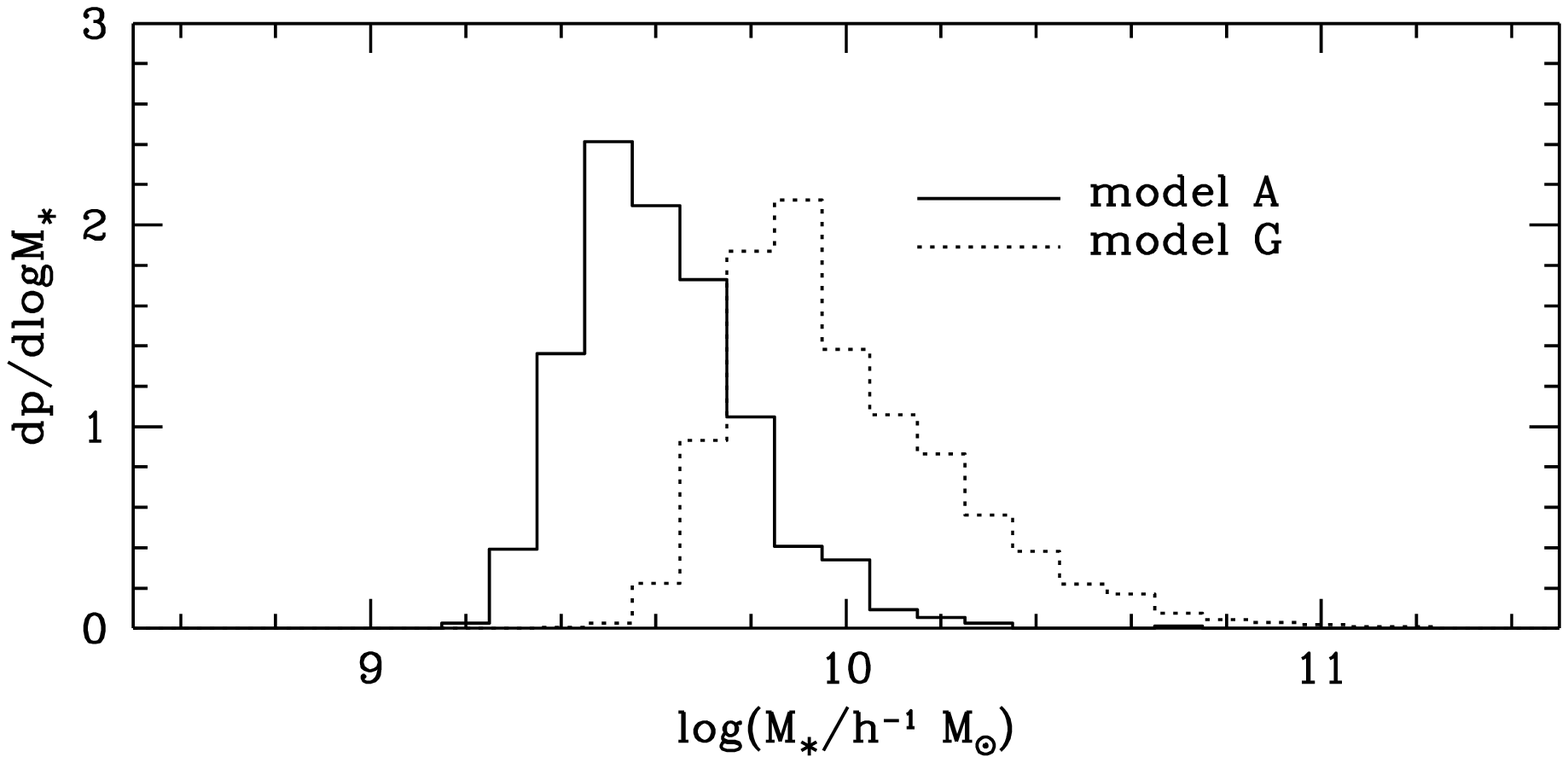}}
}
{\small {\bf FIG. 5}
The stellar mass distribution of Lyman-break galaxies. Galaxies brighter
than $\R_{\rm AB} = 25.0$ satisfying the colour criteria of
Steidel \etal (1995) are included.  The $\Omega_0=1$ model~A is shown 
by the solid line and the low-$\Omega$ model~G by the dotted line.
\label{fig:mstars}
}
\vspace{1cm}

The distribution of star formation rates in our model Lyman-break
galaxies is shown in Fig. 7a. These are instantaneous
rates, measured directly from the mass of cold gas turned into stars
per unit time. The star formation rates at $z\simeq 3$ are typically a
few solar masses per year, and are somewhat larger in the flat
low-$\Omega$ model than in the standard model. Only a very small
fraction of the galaxies at these redshifts have star formation rates
in excess of $10 h^{-2} {\rm M_{\odot} yr^{-1}}$. The instantaneous
star formation rates in typical galaxies at $z\simeq 3$ are comparable
to their mean rates averaged over the age of the universe at that
redshift (1.6 Gyr in model A, 2.5 Gyr in model~G). 
We consider the history of star
formation in more detail in Section~5.

Whereas the star formation rate of a galaxy is not directly observable, the
distribution of $\R$ magnitudes is. Since the $\R$ band samples the rest 
frame ultraviolet at $z\simeq 3$, the distribution of $\R$ magnitudes 
is closely related to 
the distribution of star formation rates. In Fig. 7b we plot
the distribution of absolute magnitude $M_{R}(\rm AB)$ in our models. This
distribution, however, is not a particularly strong constraint on the
models because, by design, the S96 survey covers only a narrow range in
$\R_{\rm AB}$. We defer a detailed comparison between our predicted star
formation rates and observations to Section~5.

\subsubsection{Galaxy sizes}

An essential feature of hierarchical models of galaxy formation is
that the typical sizes of galaxies increase with time. Thus, we expect
the characteristic radii of galaxies to be considerably smaller at
high redshift than they are at present. A detailed investigation of
the evolution of galactic sizes will be presented elsewhere (Lacey
\etal in preparation). A rough indication of the sizes of high
redshift galaxies, however, may be obtained from a simple model assuming that
galaxies acquire their angular momentum from tidal torques and that
the angular momentum of the gas is conserved as it condenses within
its halo. 
 
\centerline{
{\epsfxsize=8.truecm \epsfysize=14.truecm 
\epsfbox[0 150 580 720]{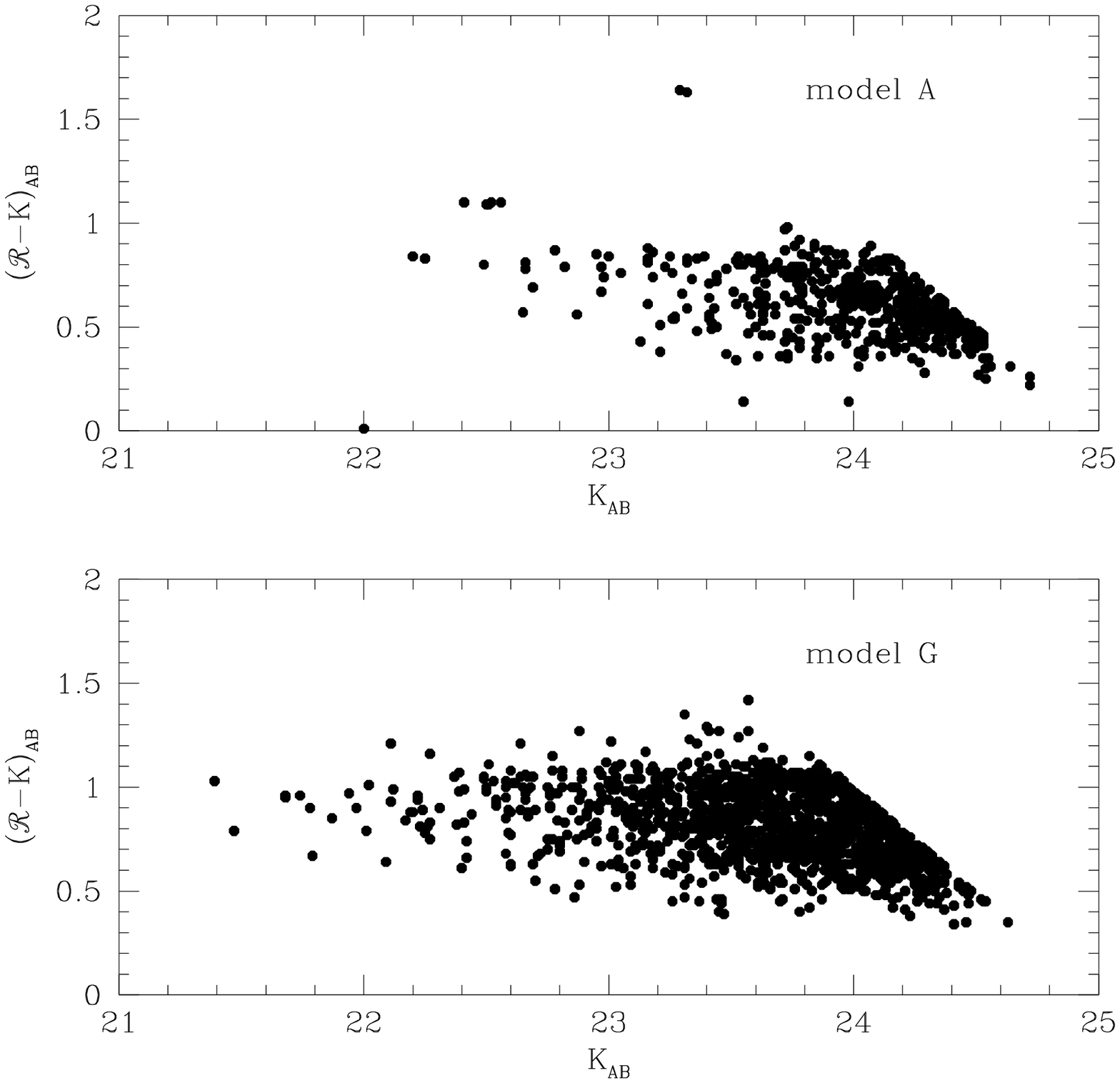}}
}
{\small {\bf FIG. 6}
The predicted $({\cal R}-K)_{AB}$ colour distribution of galaxies 
brighter than ${\cal R}_{AB} = 25.0$ satisfying the 
Steidel \etal colour selection. 
The top panel shows model A and the lower panel shows model G.
\label{fig:rk}
}
\vspace{1cm}

\centerline{
{\epsfxsize=8.truecm \epsfysize=14.truecm 
\epsfbox[0 150 580 720]{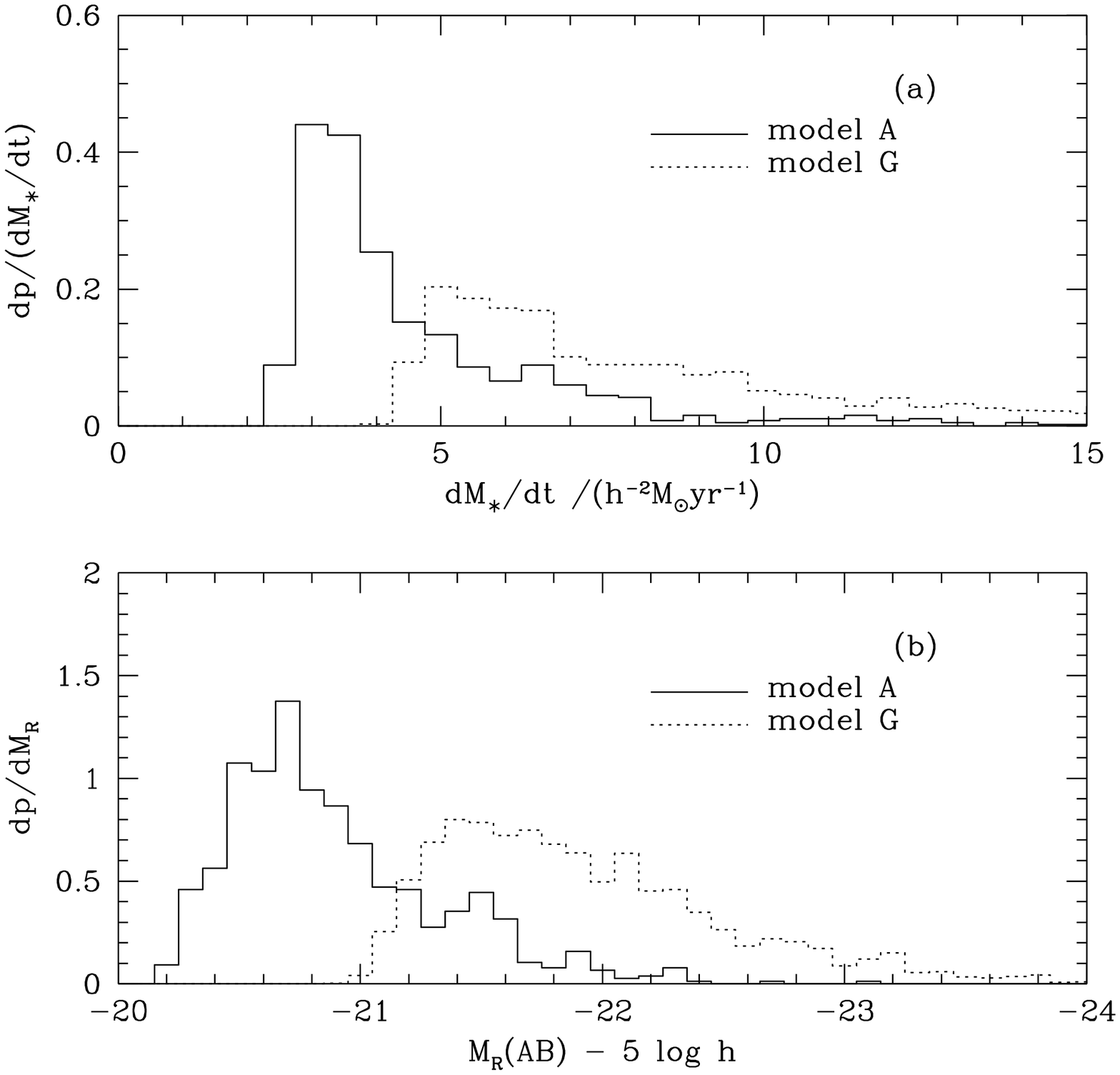}}
}
{\small {\bf FIG. 7}
Star formation rates and absolute $\R_{\rm AB}$ magnitudes for high
redshift galaxies.  Galaxies with apparent magnitudes $\R_{\rm AB} <
25.0$ satisfying the colour criteria of Steidel \etal (1995) are
included.  Model A is shown by the solid line and model~G by the
dotted line. The top panel gives the distribution of instantaneous
star formation rates in the models and the bottom panel the
distribution of absolute magnitudes, $M_\R$.  
\label{fig:sdot}
}
\vspace{1cm}

This simple model is quite adequate because, as we discuss
in Section~4, high redshift galaxies in our model tend to have very
small bulges or no bulge at all.  From this simple model, we obtain
half light radii $r_{\rm h} \sim 0.4 h^{-1} {\rm kpc}$ at $z\simeq 3$
in model A and $r_{\rm h} \sim 0.6 h^{-1} {\rm kpc}$ in model G.  This
rough calculation agrees reasonably well with the values of $r_{\rm h}
\simeq 0.7 - 1.0 h^{-1} {\rm kpc}$ ($\Omega_0=1$) or $r_{\rm h}
\simeq 1.0 - 1.5 h^{-1} {\rm kpc}$ ($\Omega_0=0.3$, $\Lambda_0=0.7$)
measured by Giavalisco, Steidel \& Macchetto (1996) from HST follow-up
observations of S96 Lyman-break galaxies.

\subsubsection{Clustering properties}

We calculate the expected clustering of high redshift galaxies in two
basic steps. First, we calculate the non-linear power spectrum $P(k,z)$ of
fluctuations in the {\it mass distribution} in comoving coordinates,
using the approximate linear to non-linear transformation of Peacock
\& Dodds (1996).
Next, we obtain a bias parameter for the galaxies using
the prescription of Mo \& White (1996). This gives the bias of dark
matter halos of mass $M$ at redshift $z$ as
\begin{equation}
b(M, z) = 1 + \frac{1}{D(z) \delta_{c}(z)} 
\left[ \frac{\delta_{c}^{2}(z)}{\sigma^{2}(M)} - 1 \right], 
\label{eq:bias}
\end{equation}
where $\sigma(M)$ is the rms linear density fluctuation at $z=0$ in a
sphere of mass $M$, $D(z)$ is the linear growth factor, and
$\delta_{c}(z)$ is the extrapolated critical linear overdensity for
collapse at redshift $z$.

\centerline{{\epsfxsize=8.truecm \epsfysize=12.truecm 
\epsfbox[0 150 580 720]{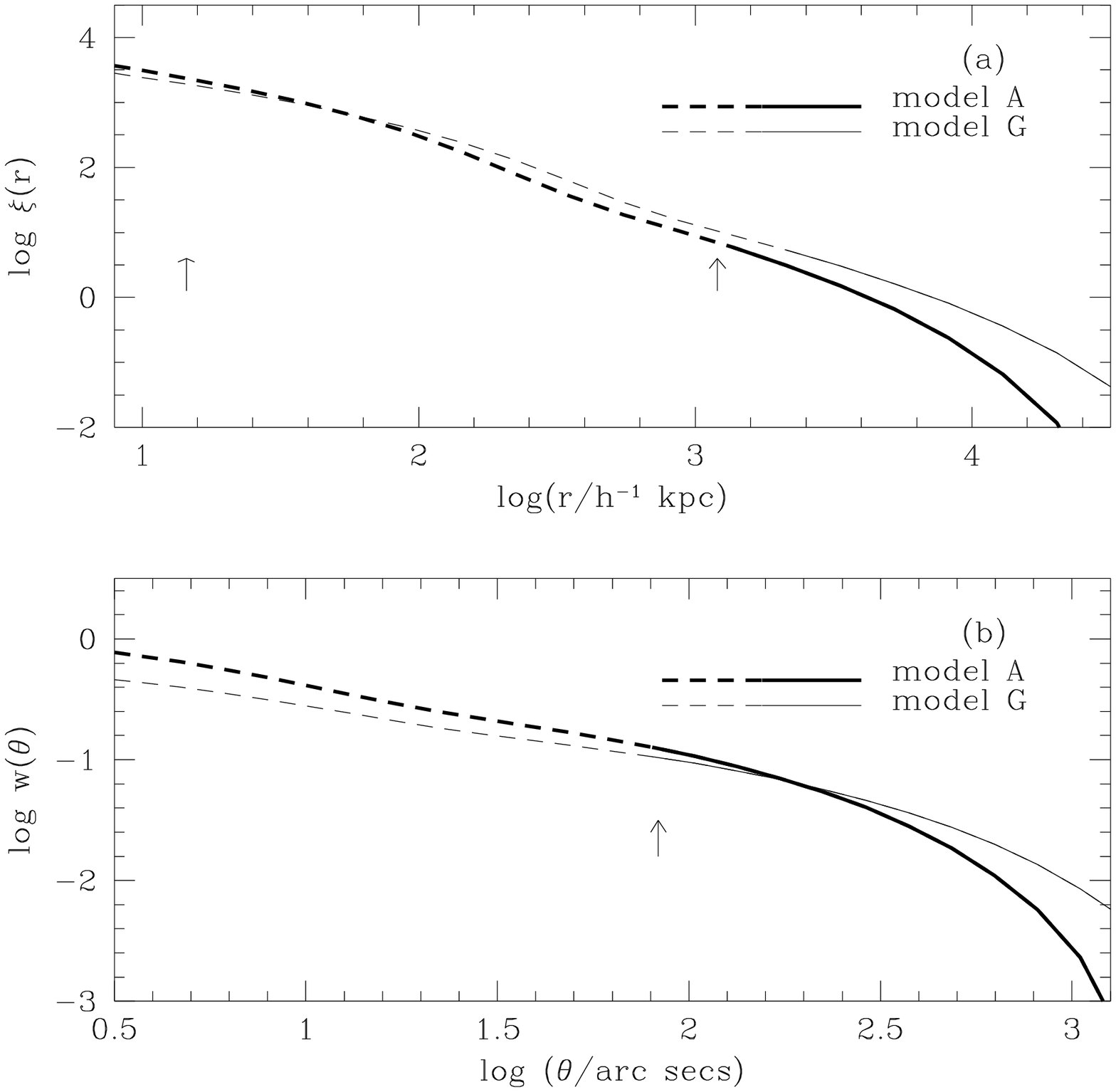}}
}
{\small {\bf FIG. 8}  
The clustering of the colour selected high redshift galaxies in
our models. Panel (a) shows the predicted spatial correlation
functions in comoving co-ordinates and panel (b) the predicted angular
correlation functions.  The heavy and light lines correspond to
correlations computed from the nonlinear power spectrum in models~A and~G
respectively, after multiplication by the bias parameter of the halos:
for model A, $\bar{b}=4.2$ and for model G, $\bar{b}=3.5$.
The solid lines show the scales at which the assumption of a constant bias 
parameter is expected to be valid.
The extrapolation of a constant bias to scales smaller than this is 
shown by broken lines. The arrow in (a) at $r = 14 h^{-1}
{\rm kpc}$ marks the comoving scale represented by 1 arcsecond at
$z=3$.  
The second arrow, at $r= 1.1 h^{-1} {\rm Mpc}$ 
indicates the comoving size of a region that collapses to form
an object with the median halo mass found for the colour selected
galaxies in model A. 
The angular scale corresponding to this comoving
length, $70$ arcseconds in model A, is shown by the arrow in (b).  
\label{fig:wtheta}
}
\vspace{1cm}

In the model, the majority of the Lyman-break galaxies are the central 
galaxy in their dark matter halo.
A good approximation to the galaxy bias function is the mean 
halo bias, $\bar{b}$, weighted
according to the mass distribution of halos that harbour galaxies
satisfying the Steidel \and (1995) colour selection criteria
(c.f. Fig. 4).  The power spectrum of the Lyman-break
galaxies is then given by $P_{\rm gal}(k,z) = \bar{b}^2 P(k,z)$. From
equation~(\ref{eq:bias}), we find $\bar{b}= 4.2$ in model A and
$\bar{b}= 3.5$ in model~G, making the approximation that the halos
all lie at the median redshift $z_{\rm m}$ of the colour selected
sample.

The spatial correlation function $\xi(r)$ of the high redshift
galaxies is obtained by taking the Fourier transform of the power
spectrum $P_{\rm gal}(k,z)$. The angular correlation function
$w(\theta)$ can be calculated using the relativistic version of
Limber's (1954) equation in the form derived by Baugh \& Efstathiou
(1993).  We make the approximation that the evolution of clustering
and the bias parameter are negligible over the narrow range of redshifts
in which Lyman-break galaxies satisfying the S96 criteria are found,
so that $P_{\rm gal}(k,z)\simeq P_{\rm gal}(k,z_{\rm m})$ in the
integral. The relativistic Limber equation can then be written as:
\begin{equation}
w(\theta) = \int k \, P_{\rm gal}(k, z_{\rm m}) \, g(k \theta) \, {\rm d} k ,
\label{eqn:wtheta}
\end{equation}
The kernel function is given by:
\begin{equation}
g(k \theta)  =  \frac{1}{2 \pi} \frac{1}{N^2} 
\int_{0}^{\infty} F(x) \left( \frac{{\rm d}N} {{\rm d} z} \right)^{2} 
\frac{ {\rm d} z}{ {\rm d} x } J_{0} (k \theta x) \, {\rm d} z, 
\label{eqn:kernel}
\end{equation}
where $x$ is the comoving distance to redshift z.  The term $F(x)$
comes from the metric and depends on the cosmology (e.g.  Peebles 1980
\S 56); in a flat universe $F(x) = 1$.  The redshift distribution
$dN/dz$ is that of the colour-selected galaxies, and $N$ is the total
number of galaxies selected.  The spatial and angular correlation
functions for galaxies satisfying the Steidel \and selection are shown
in Fig. 8a and Fig. 8b.

The derivation of the formula for the halo bias,
equation~(\ref{eq:bias}), by Mo \& White (1996) formally assumes that
the correlation function of the matter satisfies $\xi_{\rm m}(r)\lsim
1$ and that $r\gsim r_{\rm L}/2$, where $r_{\rm L} =
(3M/4\pi\rho_0)^{1/3}$ is the comoving Lagrangian radius of the halos
($\rho_0$ being the present mean density).  The model has been tested
against N-body simulations by Mo \& White and by Mo, Jing \& White
(1996), who find that the formula works quite well in practice down to
where $r\approx r_{\rm L}$, even when $\xi_{\rm m}(r)>1$.  For the
Lyman-break galaxies in model A, $r_{\rm L} \sim 1 \mpc$ (comoving),
corresponding to $\theta\sim 70 \arcsec$ for $z=3$. Coincidentally,
this is close to the scale where $\xi_m(r)=1$.  On smaller scales, the
halo correlation function should flatten relative to the matter
correlation function.  In Fig. 8 we have therefore
plotted the galaxy correlation function as a dashed curve on scales $r
< r_{\rm L}$, where the assumption of constant bias probably breaks
down. The spatial correlations in the range $0.3 \le r \le 3 h^{-1}
{\rm Mpc}$ (comoving) are a good match to a power-law, $\xi(r) =
(r_{0}/r)^{\gamma}$, with $\gamma = 1.8$ and $r_{0} = 3.9 h^{-1} {\rm
Mpc}$. Thus, our models predict that Lyman-break galaxies at $z=3$ should
have a comoving clustering length comparable to that of bright galaxies
today. 

We see that the models predict appreciable angular correlations for
the Lyman-break galaxies, $w(\theta) \approx 0.1$ at $\theta =
100\arcsec$ for model A, and a slightly lower value for model~G.  In
contrast, Brainerd \etal (1995) estimated $w(\theta) \approx 0.005$ at
the same angular scale for field galaxies with $\R_{\rm AB} \lsim
25$. The larger $w(\theta)$ that we predict is the result of several
effects: (i) the relatively narrow redshift range for the Lyman-break
objects, which reduces projection effects in $w(\theta)$; (ii) the
large degree of bias, $\bar{b} \sim 4$, which results from the
galaxies occuring in rare dark halos; and (iii) possible differences
in the $R$ magnitude scale between the S96 and Brainerd \etal
datasets.

\section{The Fate of high-z galaxies}
\label{s:fate}

\begin{figure*}
{\epsfxsize=18.truecm \epsfysize=22.truecm 
\epsfbox[0 60 580 680]{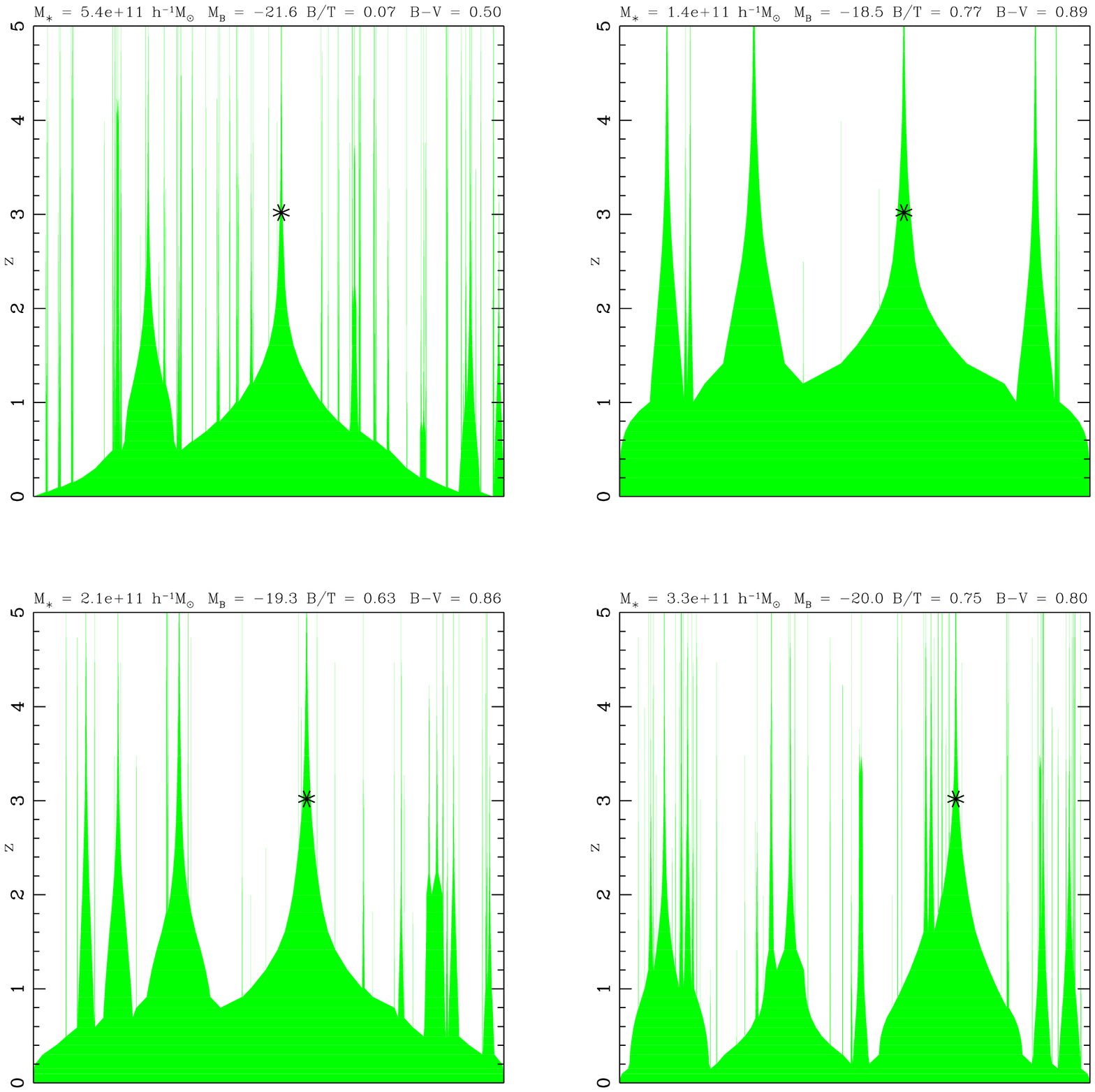}}
{\small {\bf FIG. 9}
Tree diagrams illustrating the star formation histories of four 
present day galaxies which had a high redshift progenitor satisfying the
luminosity and colour selection criteria of Steidel \and (1995). The actual
progenitor is marked by a star. 
These examples are taken from the standard CDM
model A. 
The present day is at the base of each tree - the trees extend back to a 
redshift of 5.
The width of each branch at any epoch 
is proportional to the mass in stars in the branch at that epoch.
The trees have been normalised to have unit width at $z=0.$ 
The labels give the stellar mass of the final galaxy at the present day, 
along with its B band luminosity, the bulge to total light ratio in the 
B band and the B-V colour.
\label{fig:tree1}
}
\end{figure*}

In Section~3 we showed that a population of star forming galaxies with the
abundance and global properties of the Lyman-break galaxies discovered by 
S96 arise naturally in hierarchical clustering theories. 
Models with a range of values of the cosmological parameters are 
equally successful in accounting for this population of high redshift 
galaxies. In this and the following section we consider the role that this
population plays in the general scheme of galaxy formation. First we
ask the question: what do the Lyman-break galaxies evolve into? Are they,
as S96 conjectured, the progenitors of the spheroidal components of
present day galaxies? Our semi-analytic modelling technique provides an ideal
tool to answer this sort of question since the entire evolutionary history
of a model galaxy is readily available. 

We begin by displaying graphically the evolutionary paths of a few examples.
From the present day population in the model, we have chosen four examples 
of different morphological types 
with at least one progenitor at $z \approx 3$ that
satisfied the S96 colour and magnitude selection criteria. 
The tree plots in Fig. 9 illustrate
the star formation histories of these examples. The stellar mass 
of the final, present day, galaxy is given at the top of each panel, 
along with the B-band luminosity, bulge-to-total luminosity ratio in the B
band and the B-V colour. Redshift decreases down the trees, and 
the bottom of each plot represents the present day.
Each branch in the tree represents a progenitor fragment and its width at
any epoch is proportional to the mass of stars in the progenitor at that
epoch. Branches merge together when the fragments they represent merge.
The plots have been normalised to have unit width at $z=0$. 
These tree plots are similar to those used by Baugh, Cole \& Frenk
(1996b) to illustrate the formation paths of galaxies of different
morphological types (see their figures~2 and~3.) Galaxies that undergo
major mergers at recent epochs are identified with ellipticals and S0s
whereas galaxies that have grown quiescently by protracted accretion of
cooling gas (perhaps around a bulge formed by a prior merger) are
identified with spirals. Minor mergers that do not disrupt a stellar disk 
add stars to the bulge component (see Baugh \etal 1996b for precise definitions.)

At the present day, the galaxy in the top left hand corner of Fig. 9 is a spiral, the galaxy at the bottom left is on the border between 
being an elliptical or S0 galaxy, the galaxy at the top right is a field 
elliptical and the galaxy at the bottom right is a cluster elliptical.
Lyman-break galaxies (marked by the star in each
panel) can therefore end up in galaxies of any morphological type and, as
we shall see in Fig. 11, they can span a wide range of
luminosity.

The distribution of bulge-to-total stellar mass amongst the descendants of
Lyman-break galaxies is similar to that of bright galaxies ($M_{B} - 5 \log h =
-19$) without such a progenitor. The distribution of halo circular velocity
for the Lyman-break descendants, on the other hand, is biased towards large
values typical of groups and clusters (Fig. 10). This is just
what was expected in view of the strong clustering bias exhibited by the
Lyman-break galaxies themselves (see \S3.3.4). 

The luminosity function of the present day descendants of Lyman-break
galaxies is plotted in Fig. 11 where it is compared with
the luminosity function of the galaxy population as a whole. In this figure
we show results for both the standard CDM model A and the low-$\Omega$
model~G. In both cases, the bright end of the current luminosity function
is made up of galaxies which, at $z\simeq 3$ had at least one progenitor 
that satisfied the luminosity and colour criteria required to qualify as a
Lyman-break galaxy in the study of Steidel \etal (1995). Virtually all
present day galaxies with $L \gsim 2.5 L_{\ast}$ have such a
progenitor. The fraction of Lyman-break descendants decreases with
decreasing luminosity so that virtually no present day galaxy with $L \lsim
L_{\ast}/5$ was ever a Lyman-break galaxy of the type observed by Steidel
\etal 

The assembly of the Lyman-break galaxies themselves is illustrated in
Fig. 12 where we plot the growth of the stellar mass of
selected Lyman-break galaxies with time. The stellar mass at each
redshift (which may be spread amongst several fragments) is plotted as
a fraction of 
the ``final'' stellar mass of the Lyman-break galaxy at $z=3$ in 
Fig. 12(a). 
Star formation in the Lyman-break galaxies begins very early, at $z>6$, but only
$\sim 20-40\%$ of the stars have formed by $z=4$. The bulk of the stars
present in these Lyman-break galaxies at $z\simeq 3$ was formed in the
preceeding few hundred million years.
The total star formation rates 
summed over the fragments, in units of the star formation rate 
at $z=3$, are plotted in Fig. 12(b).

{\centerline{\epsfxsize=8.8truecm \epsfysize=8.truecm 
\epsfbox[70 400 580 780]{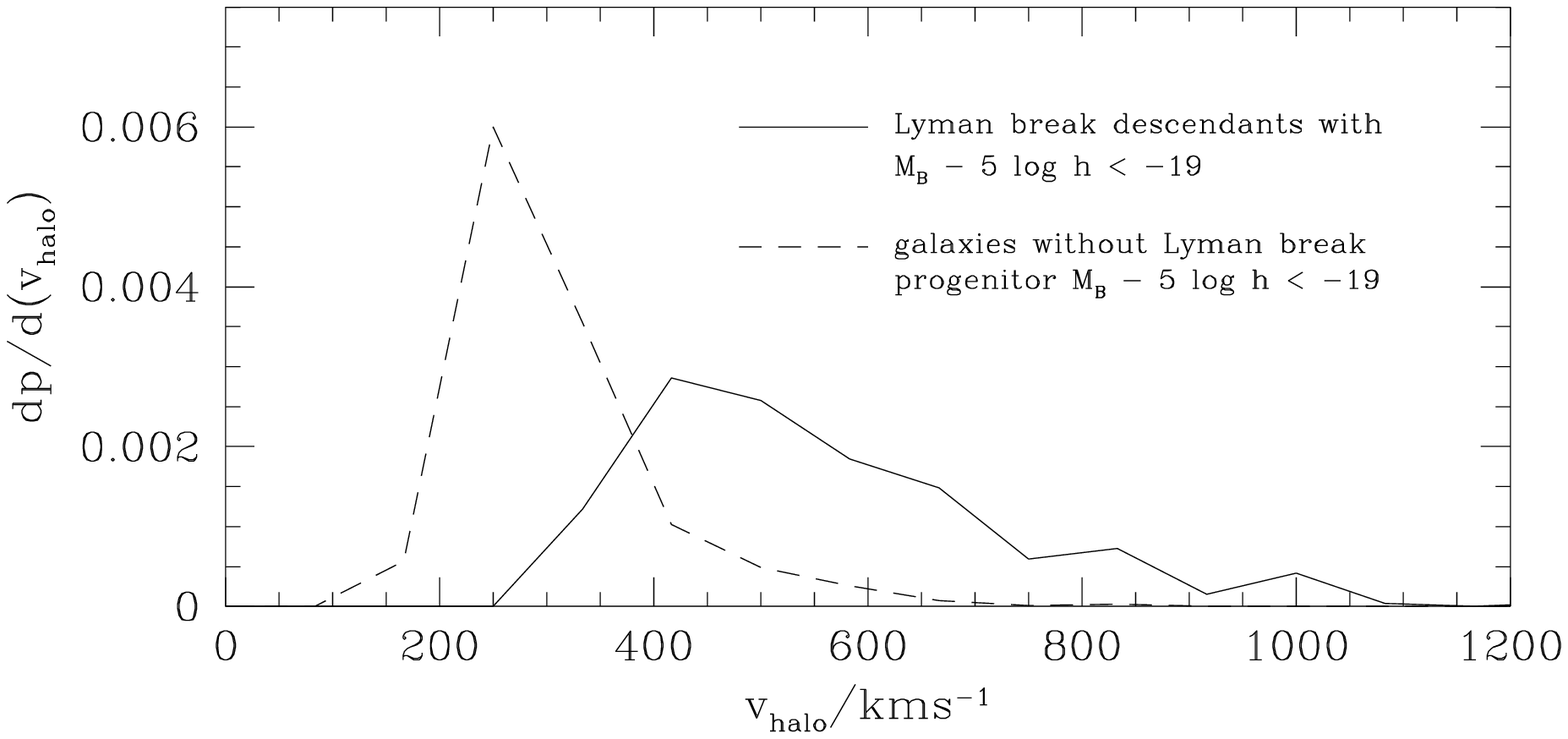}}
}{ \small {\bf FIG. 10}
The distribution of present-day halo circular velocity of galaxies in 
model A that contained at least one Lyman-break galaxy at high 
redshift (solid line), compared with the distribution of halo 
velocities for present day galaxies without such a progenitor (dashed
line). All the present day galaxies considered are brighter than 
$M_{B} - 5 \log h = -19$. 
\label{fig:vhalo}
}

\vspace{1cm}
{\centerline{\epsfxsize=8.8truecm \epsfysize=9.3truecm 
\epsfbox[0 400 580 780]{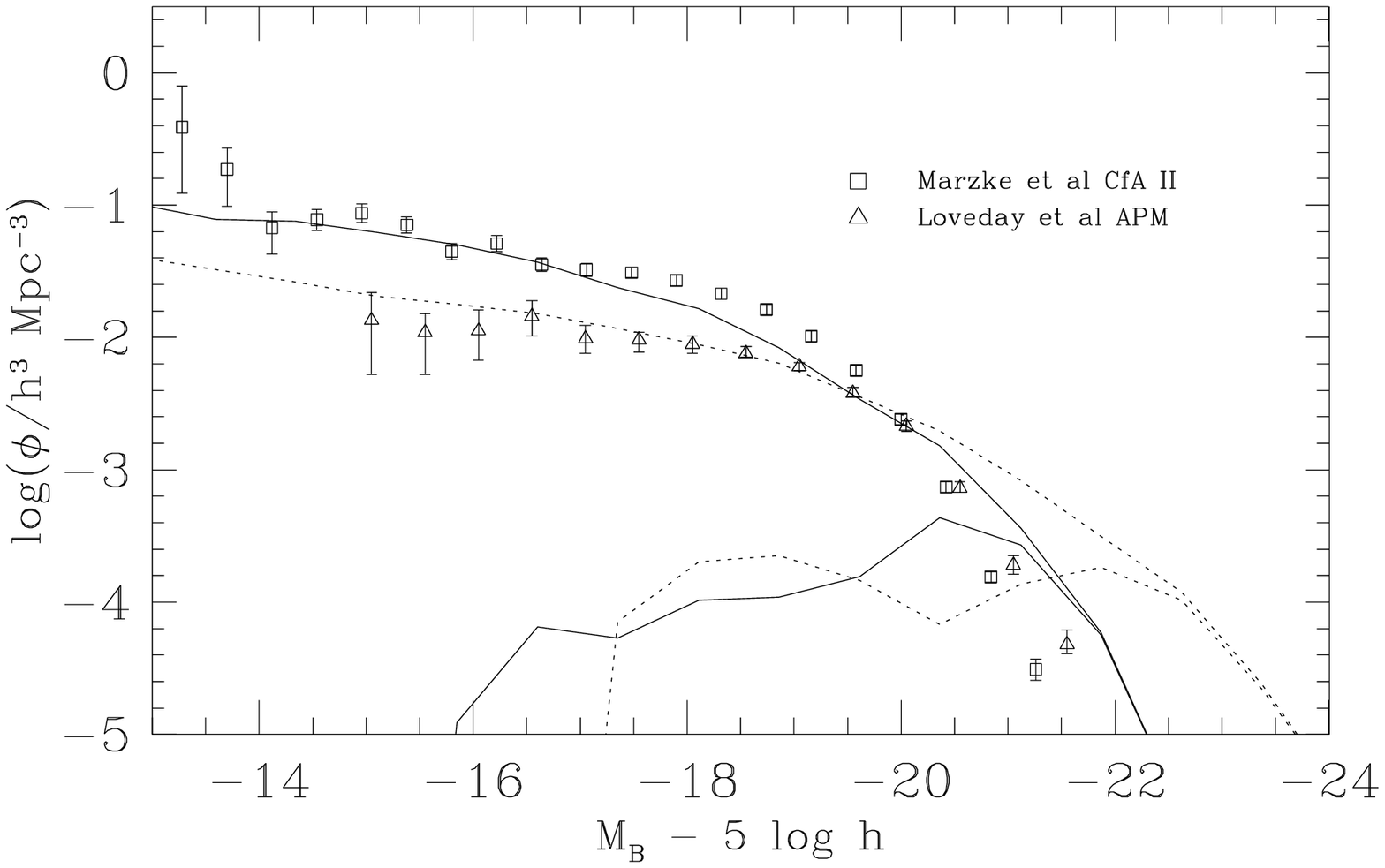}}
}
{\small {\bf FIG. 11} 
Present-day $B$-band luminosity functions in model A (solid lines)
and model~G (dotted lines). In each case, the extended curve shows the
luminosity function of the galaxy population as a whole. The shorter
curve shows the luminosity function of galaxies that contained at
least one progenitor satisfying the selection criteria for Lyman-break
galaxies at high redshift in the study of Steidel \etal (1995). The
data points show observational determinations of the luminosity
function from Loveday \etal (1992) and Marzke \etal (1994). 
\label{fig:lfdesc}
}

\vspace{1cm}
\centerline{
{\epsfxsize=8.8truecm \epsfysize=8.8truecm 
\epsfbox[0 150 580 720]{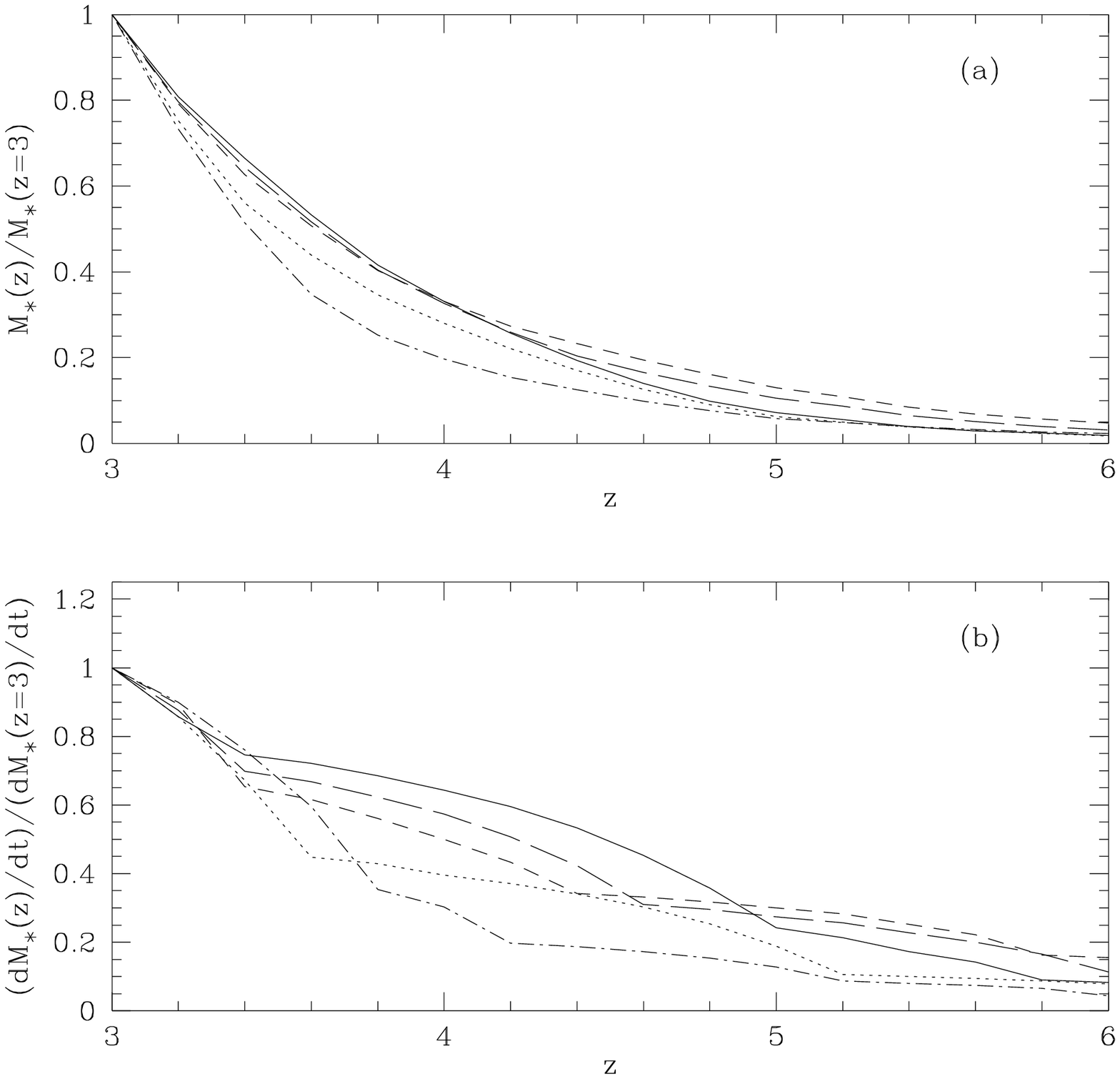}}
}
{\small {\bf FIG. 12}
The star formation histories of five example galaxies that 
satisfy the Steidel \etal colour selection in model A. The galaxies all
have dark halos of mass $2 \times 10^{12} h^{-1} M_{\odot}$ at $z=3$.
Panel (a) shows the build up in stellar mass expressed as a 
fraction of the mass in stars at $z=3$. Panel (b) shows the instantaneous 
star formation rate in units of the star formation rate at 
$z=3$. Curves of the same line style refer to the same 
galaxies in (a) and (b).
\label{fig:stars}
}

\centerline{{\epsfxsize=8.5truecm \epsfysize=8.truecm 
\epsfbox[0 150 580 720]{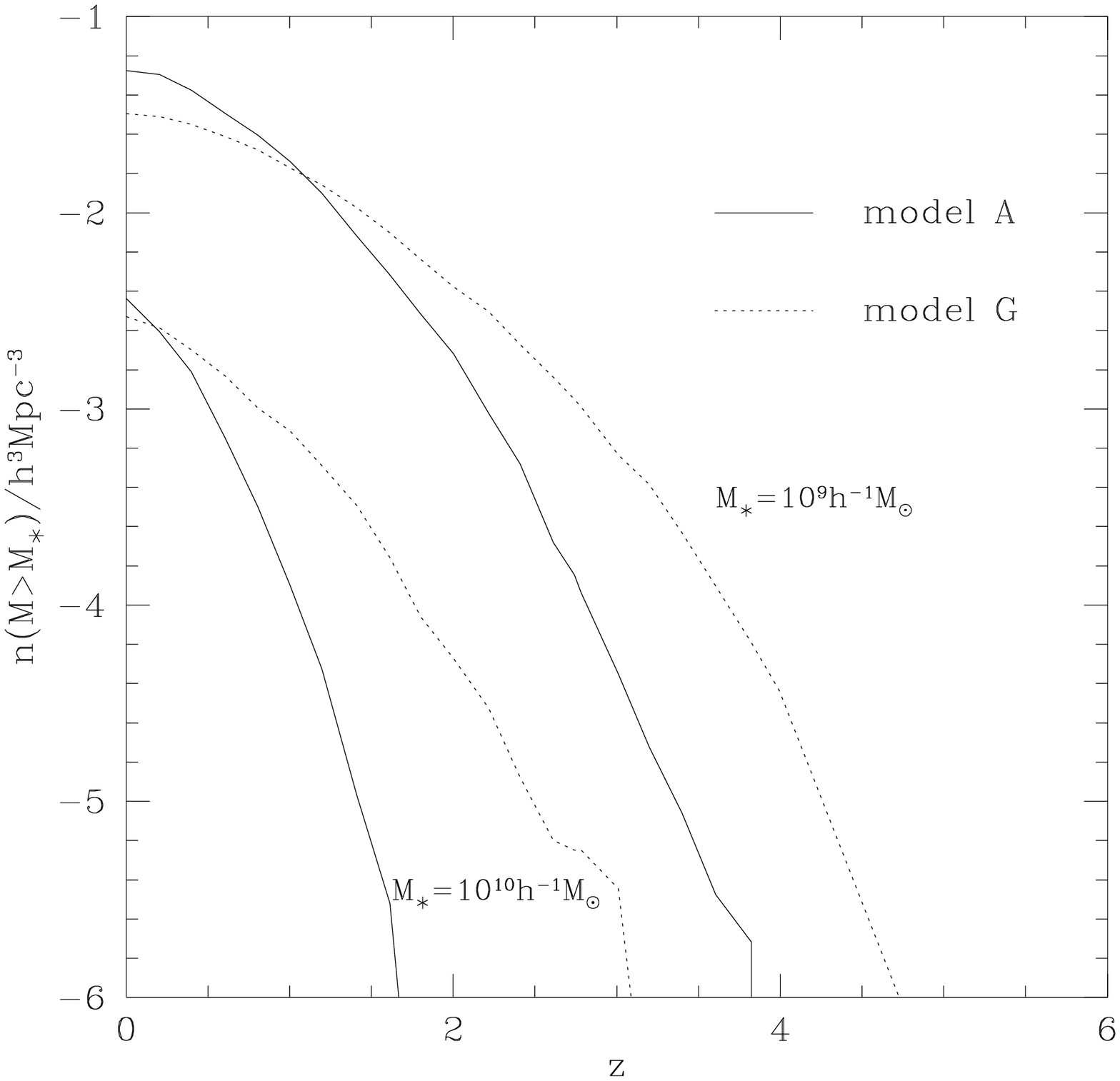}}
}
{\small {\bf FIG. 13}
Evolution of the comoving number density of galaxies that have stellar
masses in excess of $10^{9}$ and $10^{10} h^{-1} M_{\odot}$.  The solid
line shows results for model A and the dotted line for model~G.
\label{fig:mofz}
}
\vspace{1cm}

The build-up of the population of galaxies with masses typical of
Lyman-break galaxies is illustrated in Fig. 13. Here we plot
the evolution of the comoving number density of galaxies that have stellar
masses in excess of $10^{9}$ and $10^{10} h^{-1} M_{\odot}$.  The number of
galaxies above a given mass limit increases as star formation procedes, but
it decreases when mergers involving galaxies of this size occur.  At $z\simeq
3$, the abundance of galaxies with mass of a few times $10^{9} h^{-1}
M_{\odot}$ is rapidly rising. This is also close to the time when galaxies of
$M_{\star}=10^{10} h^{-1} M_{\odot}$ first appear in significant numbers.
Thus, in the cosmological models discussed in this paper, $z\simeq 3$ is 
the first epoch at which galaxies form with stellar masses comparable to 
present-day $L_*$ galaxies.

\section{The cosmic star formation history}
\label{s:epoch}

\begin{figure*}
\begin{picture}(300, 600)
\put(100, 400) 
{\epsfxsize=12.truecm \epsfysize=8.truecm 
\epsfbox[0 400 580 720]{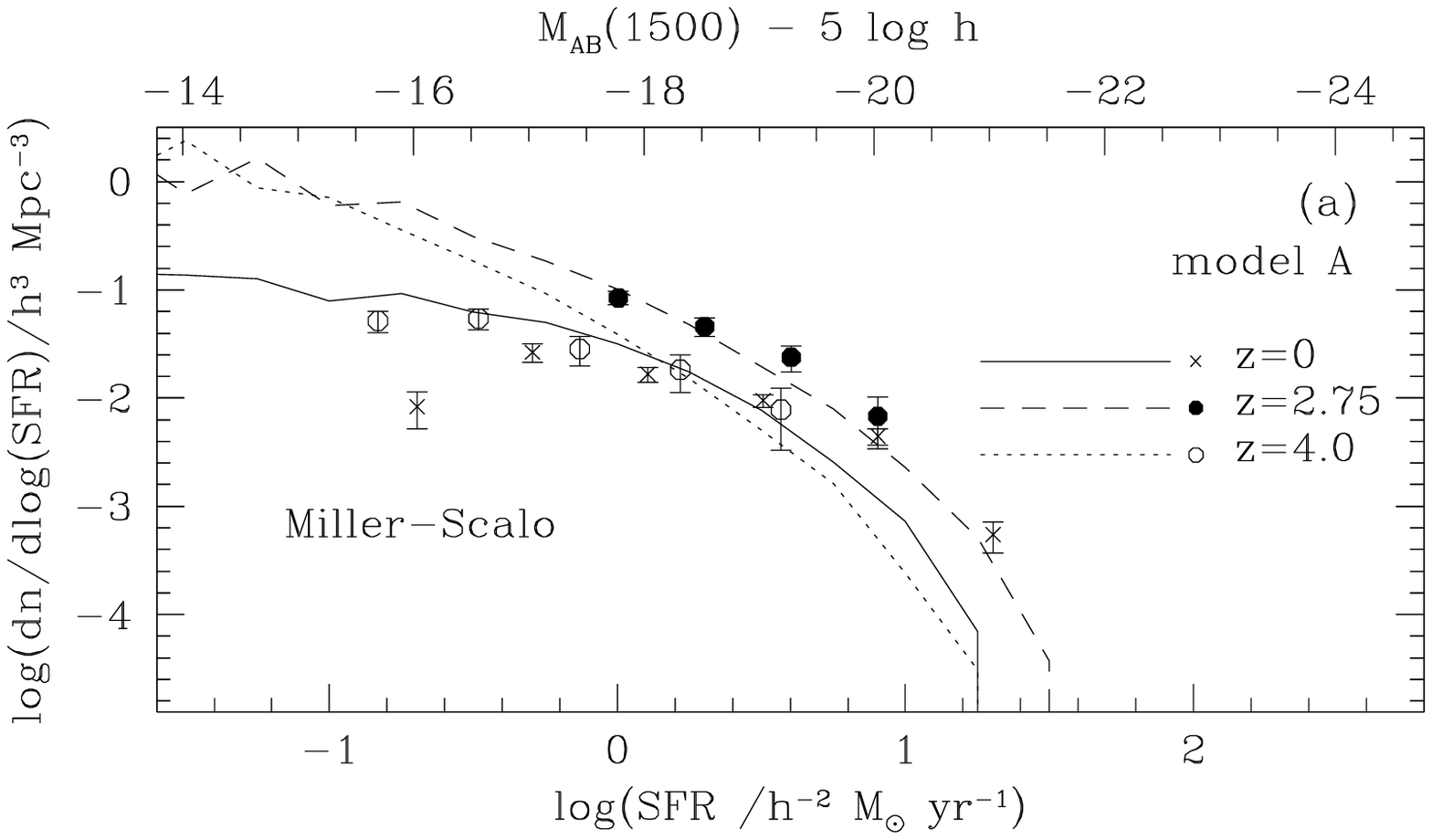}}
\put(100, 200)
{\epsfxsize=12.truecm \epsfysize=8.truecm 
\epsfbox[0 400 580 720]{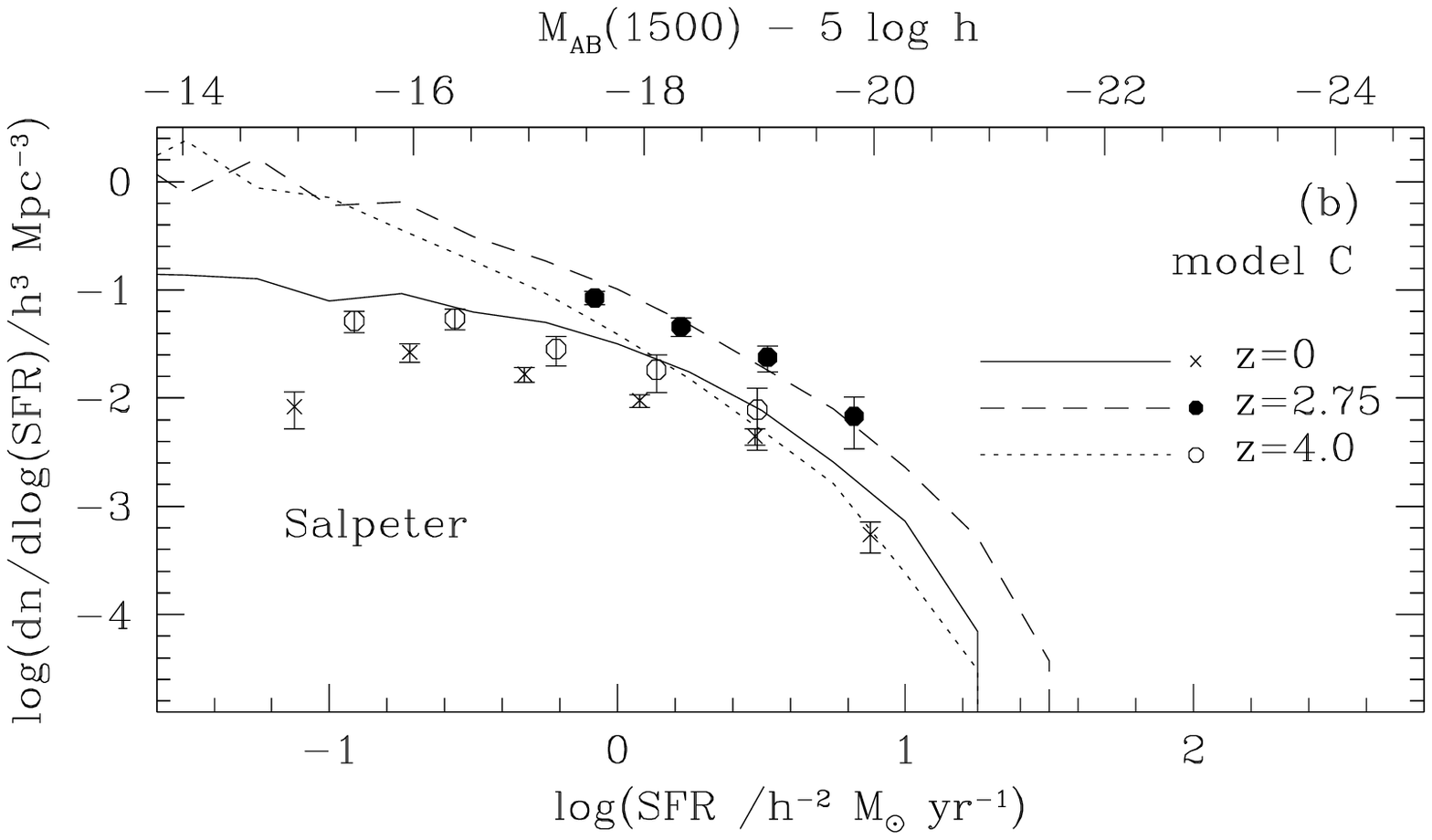}}
\put(100, 0)
{\epsfxsize=12.truecm \epsfysize=8.truecm 
\epsfbox[0 400 580 720]{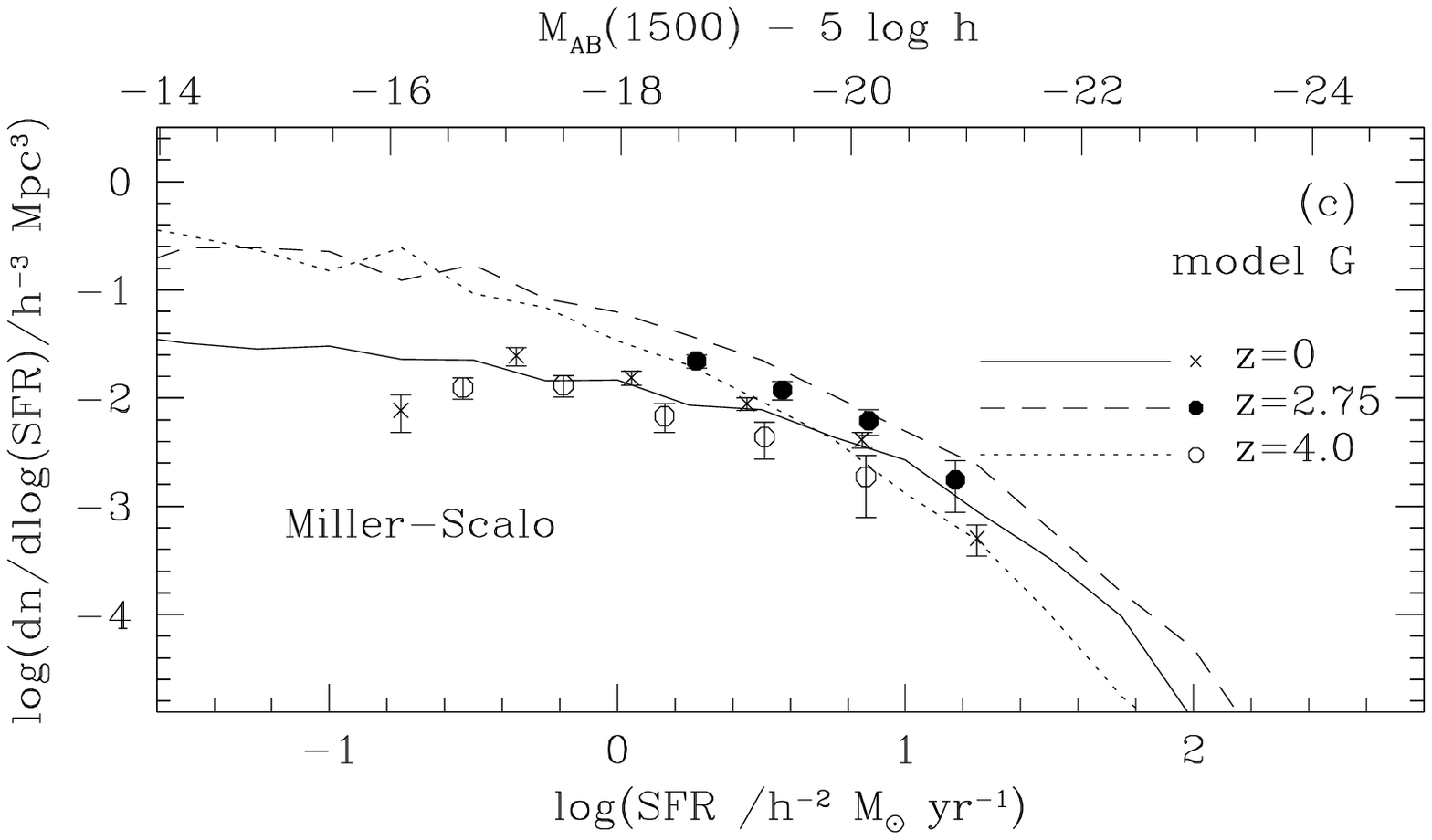}}
\end{picture}

{ \small {\bf FIG. 15}
The distribution of star formation rates (number of star forming systems
per $\log_{10}(SFR)$ per comoving volume) at different redshifts, computed
from our models and compared to observational data. Panel (a) shows model
A, panel (b) model~C and panel (c) model~G. The solid, dashed and dotted
lines show the model predictions for $z=0,2.75$ and $4$ respectively. The
symbols with error bars are observational data points. The crosses are for
$z=0$ (Gallego
\etal 1995), the filled circles for $\langle z
\rangle = 2.75$ (Madau 1996) and the open circles for $\langle z \rangle = 4$ (Madau
1996).  The observational data have been converted into total SFRs assuming
either a Miller-Scalo or Salpeter IMF, as indicated on each panel and
described in the text. The top scale shows the luminosity $L(1500)$
expressed as an AB magnitude: $M_{\rm AB}(1500) = -2.5\log
(L(1500)/\ergs\Hz^{-1}) + 51.6$.
\label{fig:sfr_z}
}
\end{figure*}

\begin{figure*}
\begin{picture}(300, 600)
\put(100, 400)
{\epsfxsize=12.truecm \epsfysize=8.truecm 
\epsfbox[40 400 580 720]{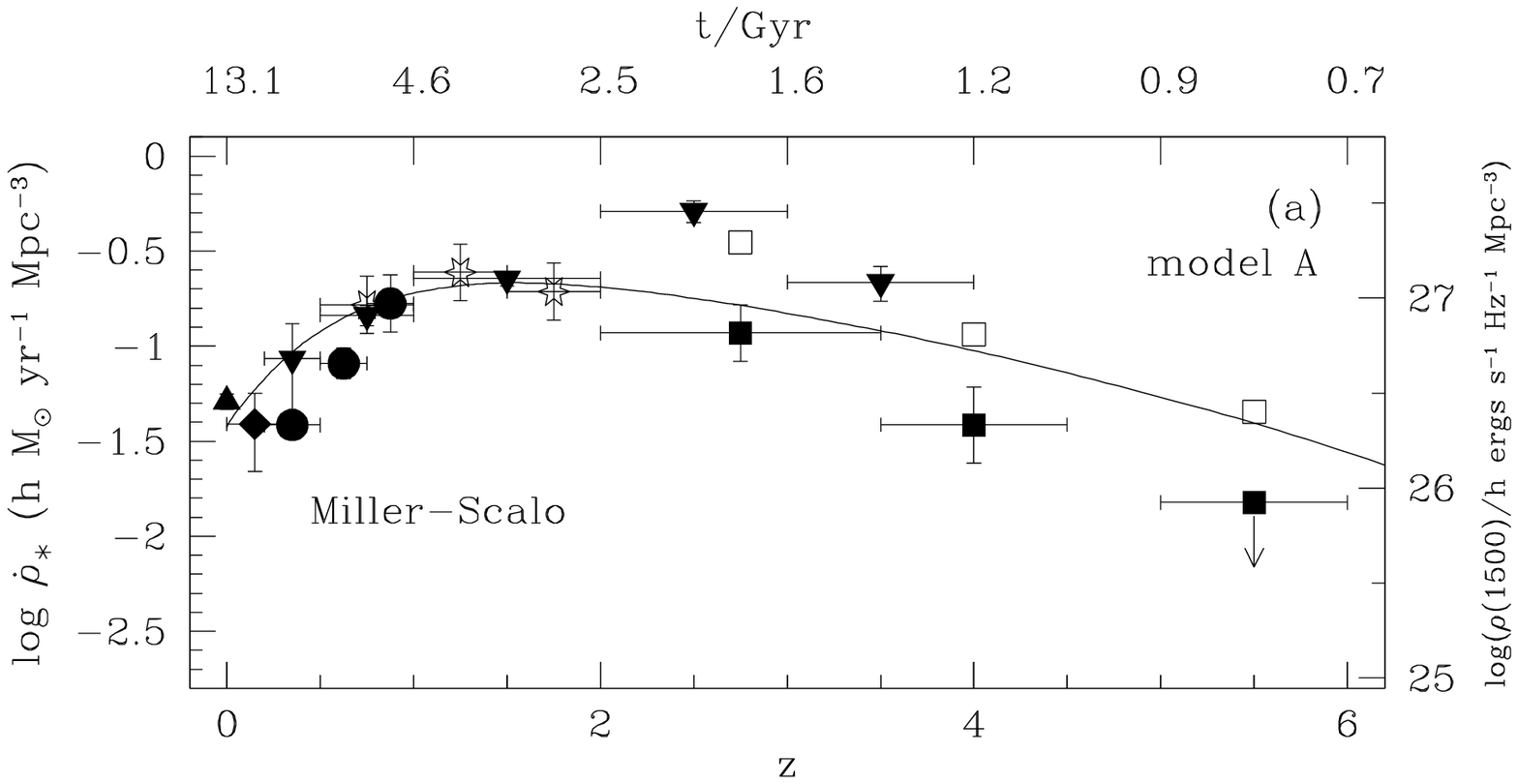}}
\put(100,200)
{\epsfxsize=12.truecm \epsfysize=8.truecm 
\epsfbox[40 400 580 720]{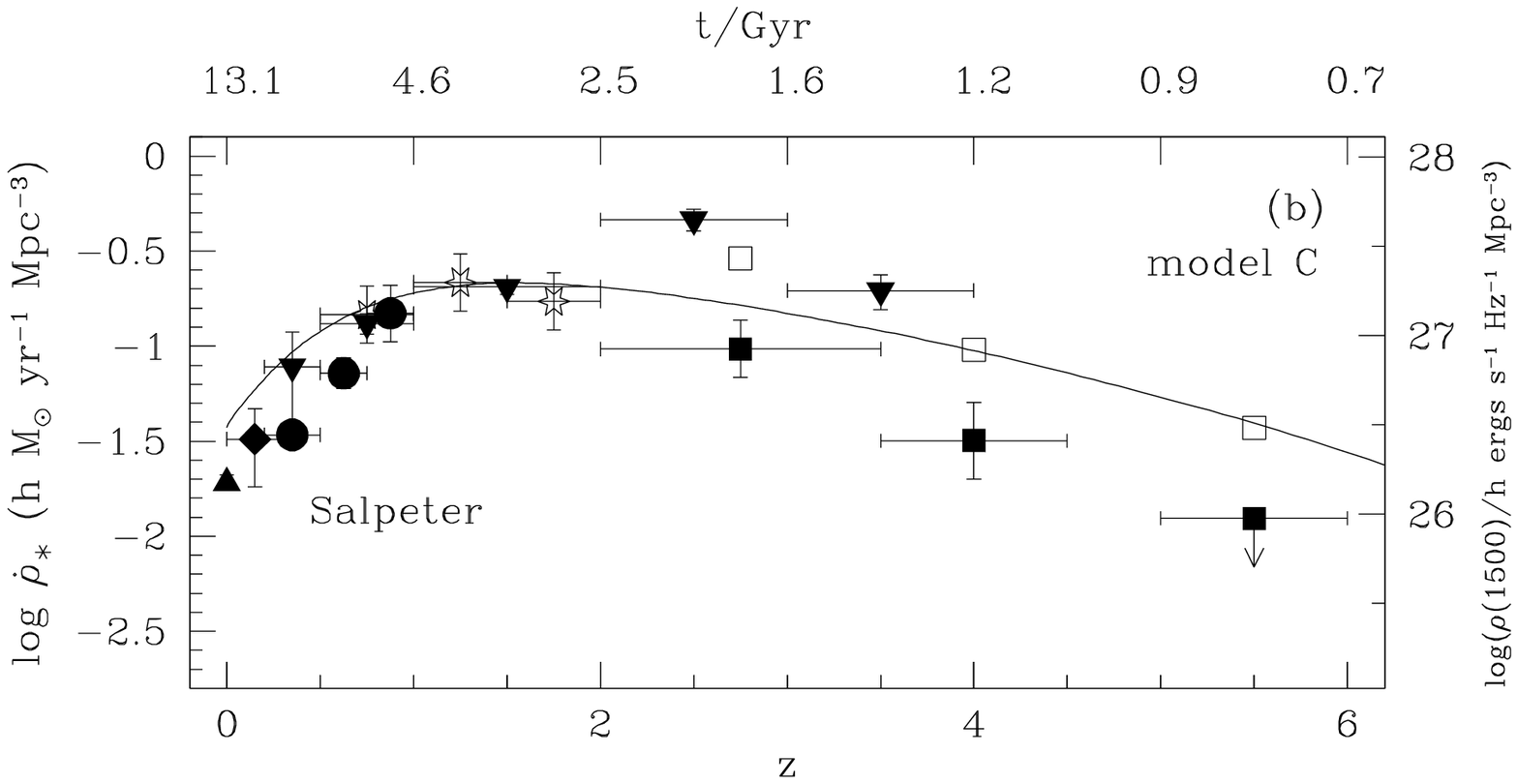}}
\put(100, 0)
{\epsfxsize=12.truecm \epsfysize=8.truecm 
\epsfbox[40 400 580 720]{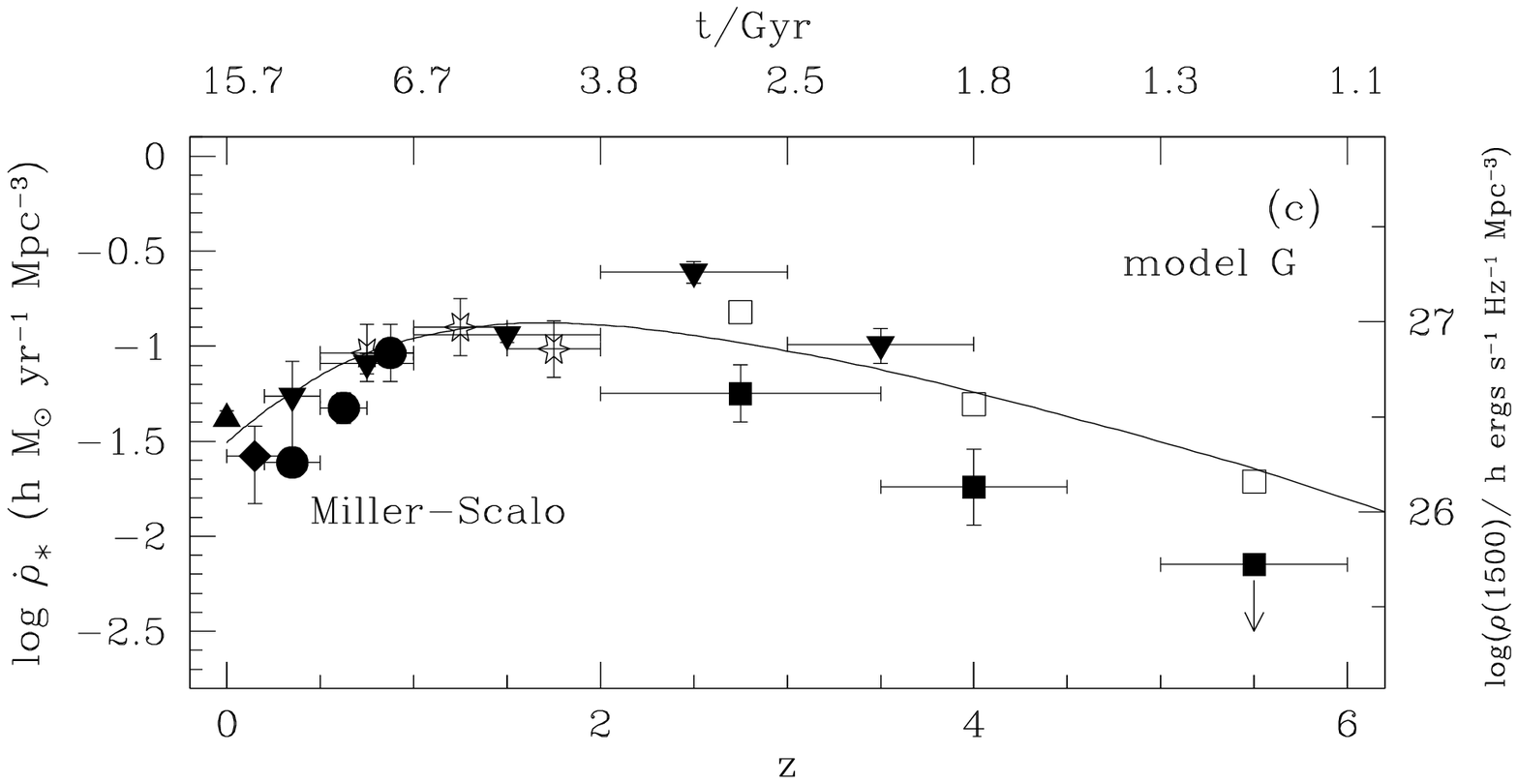}}
\end{picture}

{\small {\bf FIG. 16} 
The total star formation rate per comoving volume as a function of
redshift, computed from our models (solid curves) and compared to
observational estimates (symbols). Panel (a) is for model A, panel
(b) for model~C and panel (c) for model~G.  
The filled triangle is based on the H$\alpha$ 
luminosity function of Gallego \etal (1995); the filled diamond is 
from the $2000 {\rm \AA}$ luminosity function of 
Treyer \etal (1997); the filled circles are from the
rest frame $2800 {\rm \AA}$ luminosity densities of Lilly \etal (1996); 
the open stars come from the $2800 {\rm \AA}$ fluxes of 
Connolly \etal (1997); the inverted triangles  are from 
the $3000 {\rm \AA}$ fluxes of Sawicki \etal (1997); 
and the filled squares come from the $1500 {\rm \AA}$ luminosity 
function of Madau (1996).  
The right hand scale shows the luminosity density at $1500 {\rm \AA}$.  
The data points have been converted to total star formation rates 
assuming either a  Miller-Scalo or Salpeter IMF as indicated on each panel.
The open squares show the effects of a factor of 3 correction in Madau's
(1996), $z>2$, star formation rates per unit comoving volume due to 
dust obscuration, as suggested by Pettini \etal (1997).
\label{fig:sfr_v}
}
\end{figure*}

White \& Frenk (1991), Lacey \etal (1993), Cole \and (1994) and Heyl \etal
(1995) demonstrated that hierarchical models of galaxy formation that are
consistent with local data tend to form the bulk of their stars at
relatively low redshifts (see also Baron \& White 1987). This is a feature
not only of the $\Omega_0=1$ CDM cosmology, but also of successful
low-density CDM models. Fig. 14 shows the star formation
histories predicted in our $\Omega_0=1$ and low-$\Omega$ models (models~A
and~G, indicated by solid and dotted lines respectively).  In both
cosmologies, $50 \%$ of the stars form after $z \simeq 1$. The stars that
have formed by $z
\simeq 3$ account for less than $10 \%$ of the present day total; 
very little star formation occurs before $z = 4$. Note that in spite of the
improvements to our galaxy formation model, the curve for model~A in
Fig. 14 is virtually identical to the curve for the fiducial
model in Figure~21 of Cole \and (1994) while the curve for model~G agrees
well with the results tabulated in Table~3 of Heyl \etal (1995).

Observational data that can be compared with theoretical predictions
for the cosmic star formation history are now becoming available
(e.g. Lilly \etal 1996, Madau \etal 1996). In Fig. 15 we
present a compilation of current data expressed as the comoving number
density of galaxies as a function of star formation rate (SFR) at
different redshifts. Star formation rates are not, of course, directly
observed but inferred from the flux in a restframe UV passband, a
cosmological model to convert flux to luminosity, a model for the
spectral energy distribution, and an assumption about the initial
stellar mass function (IMF). To intercompare different datasets
amongst themselves and with our model predictions, we have derived
SFRs from published data in a homogeneous manner.
We present results for both the Miller-Scalo and Salpeter IMFs.
The SFRs in our models are total and include the
contribution from brown dwarfs (ie stars with mass below the
hydrogen-burning limit) which is parametrized by the factor
$\Upsilon>1$ (see Table~1).

\vspace{1cm}
\centerline{{\epsfxsize=8.5truecm \epsfysize=8.5truecm 
\epsfbox[0 400 580 720]{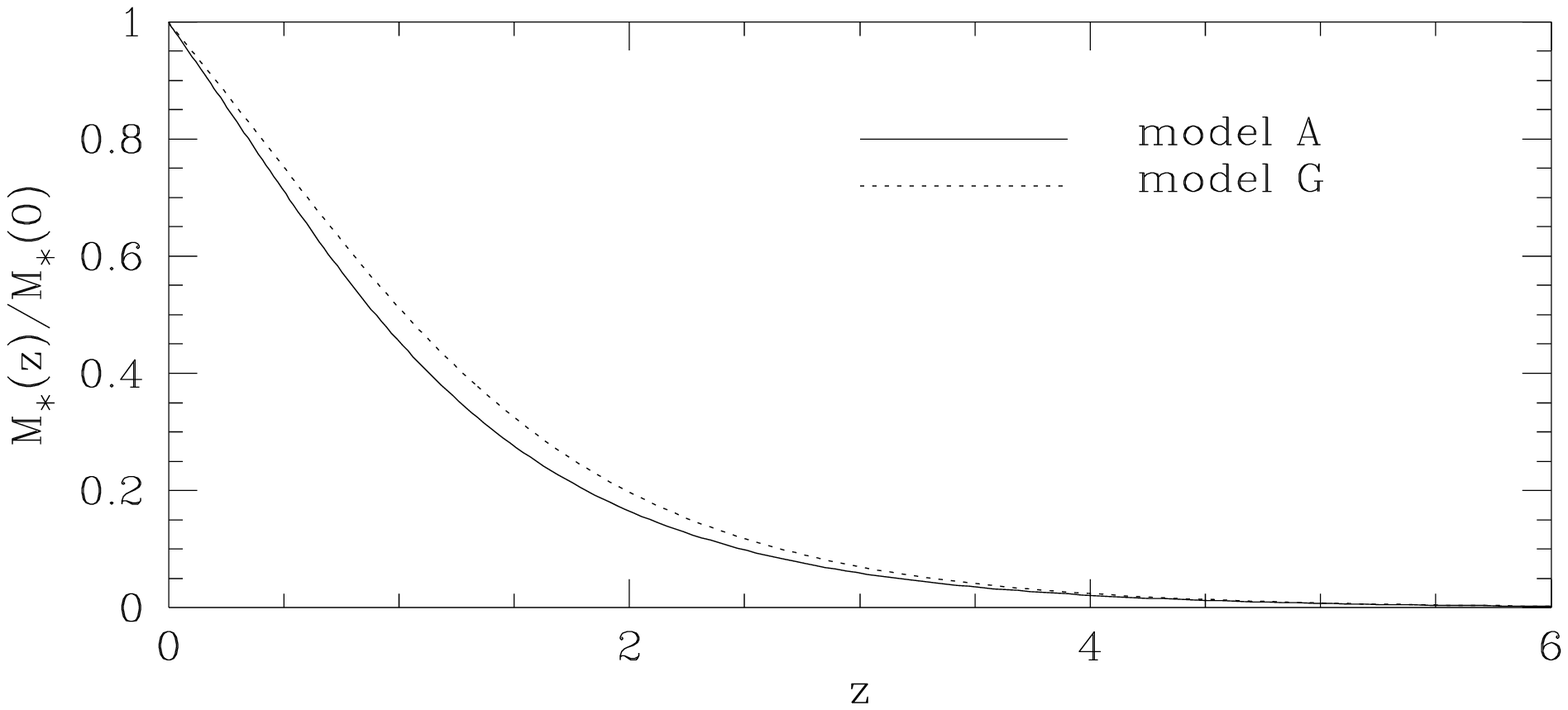}}
}
{\small {\bf FIG. 14}
Predicted star formation histories in model A (solid line) and model~G
(dotted line). The curves give the fraction of the final mass 
in stars that has formed by a given redshift.  
\label{fig:epoch}
}
\vspace{1cm}

\begin{table*}
\begin{center}
\caption[dummy]{
The $L(1500)$, $L(2800)$ and H$\alpha$ luminosities 
produced by a constant total SFR $= \Upsilon M_{\odot} {\rm yr}^{-1}$ 
after $1$ Gyr. $L(H\alpha)$  in units of ergs s$^{-1}$, $L(1500)$ and
$L(2800)$ in $\ergs\Hz^{-1}$.}
\vspace{1cm}
\begin{tabular}{lrrrr}
\hline
\multicolumn{1}{l} {band} & 
\multicolumn{1}{l} {Miller-Scalo} & 
\multicolumn{1}{l} {Scalo} & 
\multicolumn{1}{l} {Salpeter} & \\
\hline 
$L(H\alpha ) $ &  $1.38\times 10^{41}$   & $6.21\times 10^{40}$  & $1.98\times 10^{41}$  \\
$L(1500)$        &  $1.34\times 10^{28}$    & $3.88\times 10^{27}$  & $8.68\times 10^{27}$ \\
$L(2800)$        &  $1.12\times 10^{28}$    & $3.59\times 10^{27}$  & $6.76\times 10^{27}$ \\
\hline
\label{tab:factors}
\end{tabular}
\end{center}
\end{table*}

The $z=0$ data plotted in Fig. 15 were derived from the
H$\alpha$ luminosity function of Gallego \etal (1995).  The
high-redshift data in the figure were derived from the data presented
by Madau (1996) for galaxies in the Hubble Deep Field identified
photometrically as Lyman-break galaxies. The $U_{300}$-band dropouts
are estimated to have $2<z<3.5$ and $\langle z\rangle = 2.75$, while the
$B_{450}$-band dropouts are estimated to have $3.5<z<4.5$ and $\langle
z\rangle = 4$.  Madau gives SFRs derived from broad-band magnitudes at
wavelengths close to $1500 {\rm \AA}$ in the rest frame.  We first
convert these back to the corresponding values of $L(1500)\equiv
L_{\nu}(1500{\rm \AA})$ using Madau's own conversion factor, as given
in Madau \etal (1996). We then convert the $1500 {\rm \AA}$ and H$\alpha$
luminosities into total SFRs using the values tabulated in
Table 5 and the appropriate value of $\Upsilon$ for
each model. In addition, for models with $\Omega_0 \ne 1$ an
approximate scaling has been applied to Madau's data points to account
for the differences in the comoving volume element and luminosity
distance in the different cosmologies.

The curves plotted on Fig. 15 display our model predictions.
The upper panel shows results for our standard $\Omega_0=1$ model~A and the
lower panel for our flat low-$\Omega$ model~G, both of which have the
Miller-Scalo IMF. The middle panel shows model~C which has $\Omega_0=1$ and
the Salpeter rather than the Miller-Scalo IMF. The theoretical curves are
exactly the same for this case as for model~A, the only difference being
the scalings applied to the observational data points. In all cases the
behaviour of the models is qualitatively the same as the observed data. The
SFRs are larger at the intermediate redshift $z=2.75$ and at $z=4$ they
drop back to values similar to those at $z=0$. The model that best
reproduces the observed data is model~A.  There are, however, considerable
uncertainties in this comparison. For example, the mean redshift of the
$U_{300}$ dropouts in our mock HST catalogue is $z\simeq 2.3$, smaller than
the central value, $\langle z \rangle=2.75$, assumed for the real data.  This
approximation alone would lead to an overestimate of the inferred star
formation rate by up to $15$\%. For the $B_{450}$ dropouts this effect is
smaller, around $5\%$. More importantly, these comparisons are sensitive to
the details of the adopted IMF, as may be seen by the way in which the 
data points shift in panels $a$ and $b$. 

The cosmic star formation history may be conveniently summarised by
considering the variation with redshift of the total SFR per comoving
volume, as in the theoretical predictions of Fig. 14.  This
quantity is obtained by integrating the differential distributions in
Fig. 15 over galaxies of all SFRs.  For the theoretical
models, this is straightforward, but for the observations, it involves
extrapolating the distribution of SFRs to ranges not directly observed.
Comparing theoretical and ``observed" total SFRs is thus considerably more
uncertain than comparing differential distributions. With this caveat, we
compare in Fig. 16 our theoretical predictions with several
observational estimates including those based on the data of Gallego \etal
(1995) and Madau (1996) already described, and also the $B$-band data of
Lilly \etal (1996) for $0.2<z<1$. In all cases, we have used the observers'
estimates of the total luminosity density, which are based on fitting a
Schechter function to the luminosity function and extrapolating it to all
luminosities. The Lilly \etal data refer to the luminosity density at $2800
{\rm \AA}$. This is less directly related to the instantaneous star
formation rate than the H$\alpha$ or $1500 {\rm \AA}$ luminosities, because
it is dominated by somewhat older stars.  The $2800 {\rm \AA}$ rest-frame
luminosity is sampled by the observed B-band flux only at $z \sim 0.5-1.0$,
so the estimate at $z \sim 0.35$ requires a modest extrapolation from 
longer wavelengths.  This introduces an additional uncertainty. The
constants used to convert $L(2800)$ to total SFR are also listed in
Table 5.  The upper limit plotted at $z=5.5$ is based on
the number of $V_{606}$ dropouts in the HDF, which are candidates to be
Lyman-break galaxies at $5<z<6$ (Madau, private communication).

After the original version of this paper was submitted, many more data
points have been added to the star formation history diagram and we
reproduce here a selection of them. At low redshift, Treyer
\etal (1997) have estimated the star formation density from 
$2000 {\rm \AA}$ (rest-frame) 
fluxes; Sawicki \etal (1997) have used photometric
redshifts of galaxies in the HDF to infer the star formation density from
$3000 {\rm \AA}$ fluxes; Connolly \etal (1997) 
have used both optical and ground
based near-infrared imaging of the HDF to infer star formation rates from
$2800 {\rm \AA}$ fluxes. 
(The use of infrared data is particularly important for
the accuracy of photometric redshifts at $z \sim 1-2$.)  Apart from the
Sawicki \etal points at $z>2$, the level of agreement amongst these
different determinations is remarkable, although it is suggestive that both the
Sawicki \etal and Connolly \etal points at $z \le 1$ lie above those 
from the CFRS survey, in better agreement with our model predictions. 
(The CFRS survey, however, has more galaxies and therefore smaller error 
bars at these redshifts.)

Overall, the agreement between theoretical predictions and data in
Fig. 16 is impressive. It must be borne in mind that these
are genuine theoretical predictions that predate the observational data.
The theoretical curve in the upper panel of Fig. 16 is
simply the time derivative of the integrated curve for the fiducial
CDM model plotted in figure~21 of Cole \etal (1994).  Note, however,
that the location of the observational data points depends on the
assumed IMF; Cole \etal used a Scalo IMF whereas in \S 3.2, we found a
Miller-Scalo or Salpeter IMF to be preferable. For a given luminous star
formation rate, $SFR/\Upsilon$, a Miller-Scalo IMF gives $2.2$ times
the H$\alpha$ flux, $3.5$ times the $1500 {\rm \AA}$ flux and $3.1$
times the $2800 {\rm \AA}$ flux compared to a Scalo IMF. Comparing
panels (a) and (b) of Fig. 16, we see that the main
effect of changing the IMF from Miller-Scalo to Salpeter is to move
the $z=0$ data point based on the H$\alpha$ luminosity. Both
models~A and~G show the same qualitative trend as the data, with a
broad peak in the total SFR at $1\lsim z \lsim 2$. For $z\gsim 2$, the
model SFRs fall off somewhat more slowly than the data. The
completeness of the observational data at these high redshifts,
however, is difficult to establish. 

A further source of uncertainty is the possible presence of dust in the
star-forming galaxies. Even a modest amount of dust would cause attenuation
of the ultraviolet flux, leading to a potentially severe underestimate of
the star formation rate. Tentative detections of the cosmic infrared
background by Puget \etal (1996) and Guiderdoni \etal (1997a) and upper
limits on it (Kashlinsky, Mather \& Odenwald 1996) provide only weak
constraints on the amount of dust present in high redshift galaxies (Madau,
Pozzetti \& Dickinson, 1997; Guiderdoni \etal 1997b). Monolithic collapse
models (e.g. Eggen, Lynden-Bell \& Sandage 1962) in which a significant
fraction of the total star formation in the universe takes place at high
redshifts enshrouded in dust (e.g. Meurer \etal 1997), appear to
overpredict the total mass of heavy elements in place at early times (Madau
\etal 1997), as inferred from observations of damped Lyman-$\alpha$ systems
(Pettini, Smith, King \& Hunstead 1997).  Estimates of the factor by which
star formation rates deduced from UV flux should be revised to account for
the presence of dust span a range of values.  The results are sensitive
both to the form of the extinction law adopted and to the assumed age of
the primeval galaxy which determines how intrinsically blue it is.
Primeval galaxies in our models are not, in general, ultraluminous
starbursts since they never experience exceptional star formation rates. In
this case, Dickinson \etal (in preparation) and Pettini \etal (1997) argue
that the likely correction is around a factor of $1.8-3$ at $z \sim 3$ for
star formation rates inferred from $1500 {\rm \AA}$ fluxes. These
corrections are a factor of $\sim 5$ smaller than those advocated by Meurer
\etal (1997). In Fig. 16, we illustrate the effect of
the Pettini \etal correction at high redshift by multiplying the points of
Madau \etal (1996) by a factor of 3 (open squares).  The correction
appropriate to the lower redshift points is also uncertain and we do not
attempt to illustrate it in Fig. 16. It is likely to be 
smaller than at high redshift since the star formation rates are derived
from longer wavelength data.

A related observational constraint on the evolution of the galaxy
population comes from observations of neutral hydrogen at high redshift
using quasar absorption lines. Fig. 17 compares the evolution
of the cold gas fraction in our models with estimates by Storrie-Lombardi
\etal (1996), derived from the statistics of damped Lyman-alpha absorption
lines. Whereas Kauffmann's (1996b) semi-analytic models agree quite
well with these data, our own 
models agree only in the qualitative sense that the 
comoving cold gas density has a broad peak at a redshift
$z=2$--$3$. Our models predict consistently more cold gas than is
inferred from the observations. 
On the other hand, the observational results may underestimate the
total cold gas density in galaxies because (i) they only include atomic
hydrogen at column densities $N_H > 2\times 10^{20}\cm^{-2}$, and do not
include ionized or molecular gas at all; and (ii) dust obscuration may
cause some absorption systems to be missed.
Regarding (i), all the gas in our model galaxies at $T\lsim 10^4\K$ is 
counted as ``cold''; the correction for ionized gas and for low 
column-density HI ($N_H < 2\times 10^{20}\cm^{-2}$) is probably 
not large, but the correction for molecular hydrogen might be 
significant. Regarding (ii), the chemical evolution models of Pei \& 
Fall (1995) suggest that because of dust obscuration, the
true neutral hydrogen density is 2-3 times higher than the ``directly
measured'' value, moving the observational points in Fig. 17
much closer to the theoretical curves. The dotted-line set of 
errorbars in  Fig. 17(a) show plausible corrections for these 
effects, using the output from one of the models 
of Pei \& Fall (1995), following Figure 2 of Storrie-Lombardi \etal (1996). 

\centerline{
{\epsfxsize=9.truecm \epsfysize=9.truecm 
\epsfbox{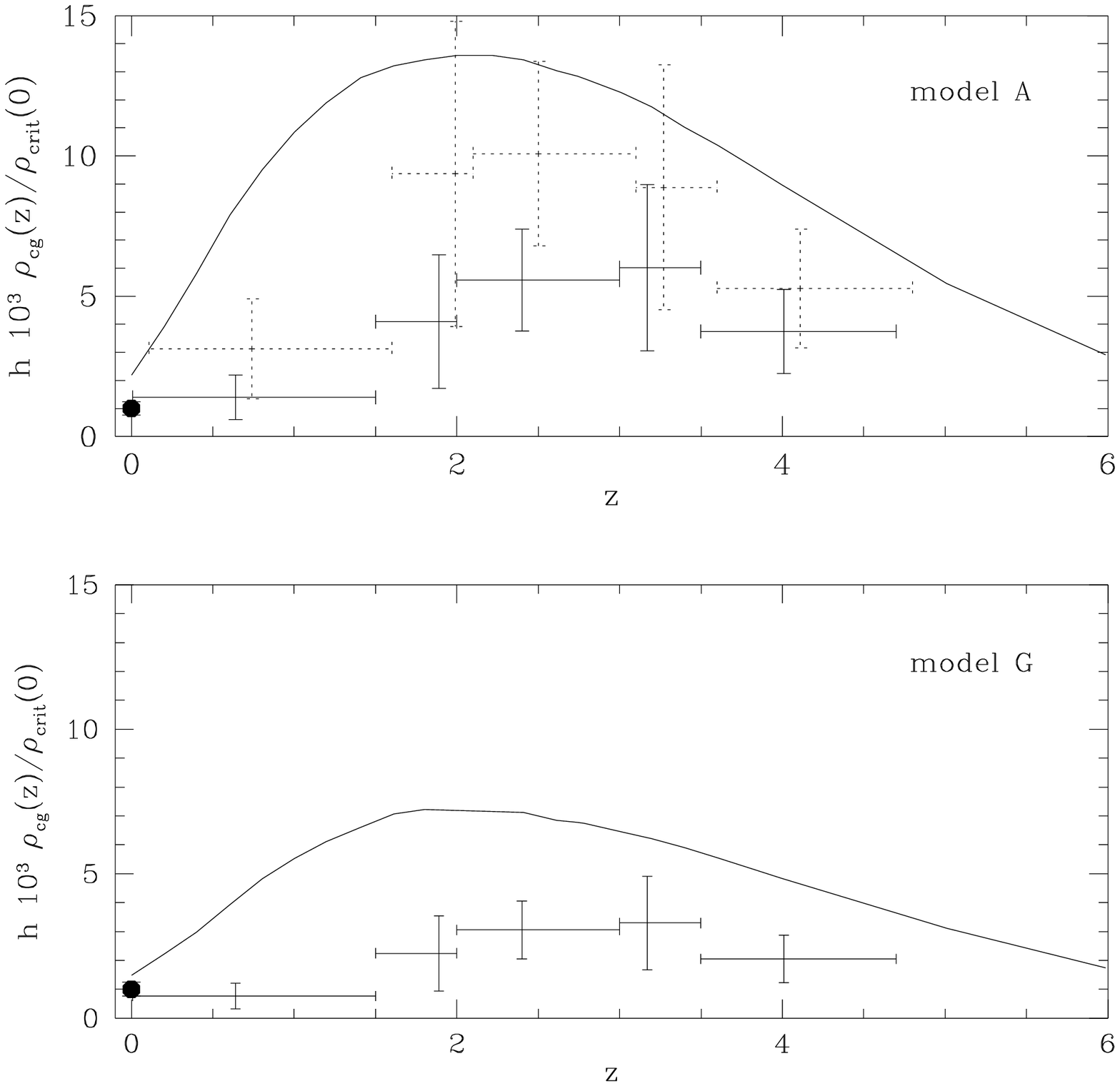}}
}
{\small {\bf FIG. 17} 
The total comoving density of cold gas, $\rho_{\rm cg}$, in units of the
present critical density.  The curve in the upper panel shows the evolution
of $\rho_{\rm cg}$ with redshift in model~A. The dependence of $\rho_{\rm
cg}$ with redshift in models B and~C which differ from~A only in the choice
of IMF, are identical to that in A. The curve in the lower panel shows
results for model~G. The data points are observational estimates, 
based on damped Lyman-alpha absorption lines from Storrie-Lombardi \etal
(1996). 
We have applied an approximate scaling to their $\Omega=1$
estimates to derive the corresponding values for the flat low-$\Omega$
model~G. The data point at $z=0$ is based on the $HI$ luminosity function
of nearby galaxies derived from $21 {\rm cm}$ observations.
The dotted-line errorbars in the upper panel show the corrections to 
the data suggested by Storrie-Lombardi \etal to account for the 
effects of incompleteness and of dust obscuration, using a model 
from Pei \& Fall (1995). We have retained 
the same fractional errors on the 'corrected' data points. 
\label{fig:cgas}
}

\centerline{
{\epsfxsize=9.truecm \epsfysize=9.truecm 
\epsfbox[30 450 580 720]{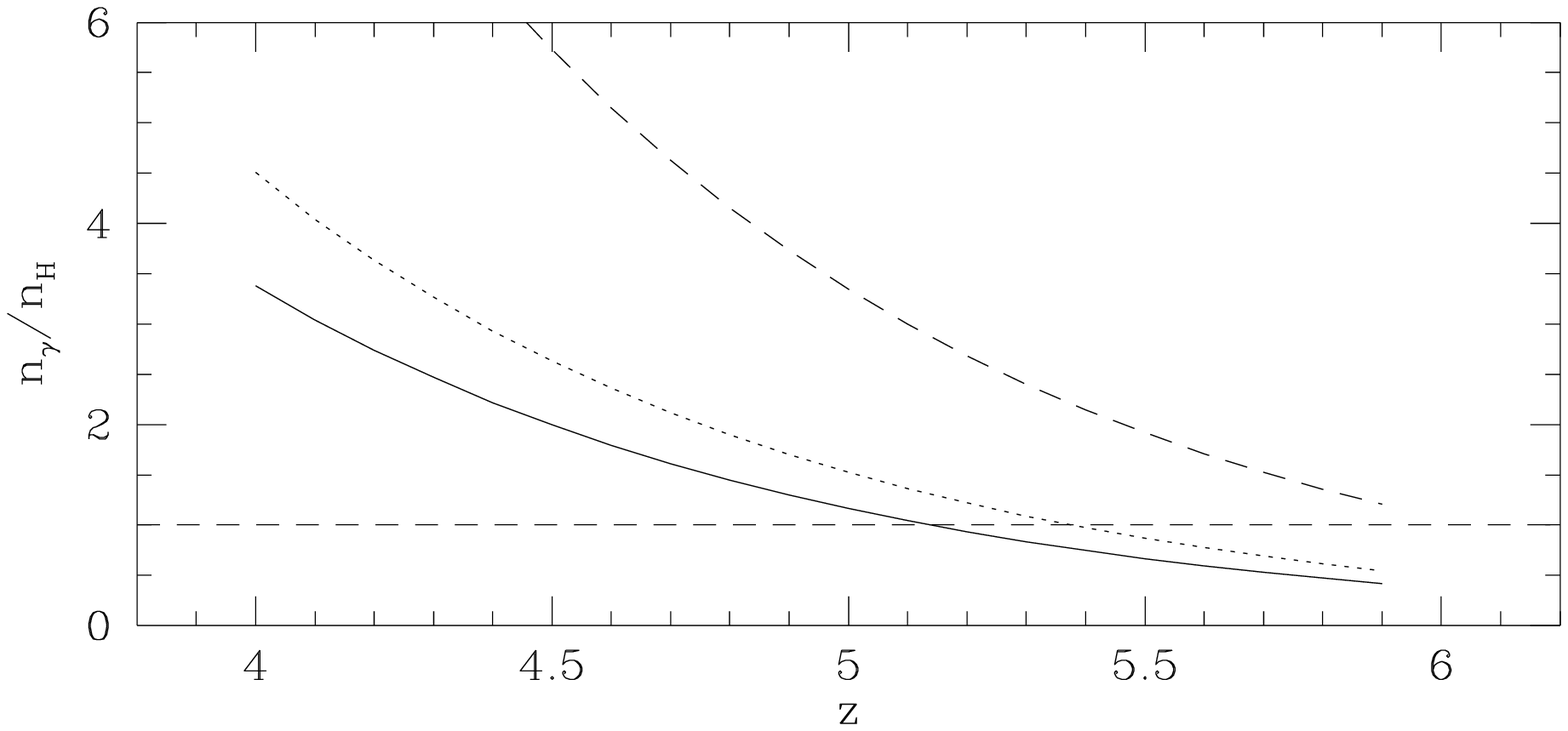}}
}
{ \small {\bf FIG. 18}
The ratio of the number density $n_{\gamma}$ of ionizing photons
produced per comoving volume to the number density $n_H$ of hydrogen
atoms. At the redshift at which $n_{\gamma}/n_{\rm H}=1$, just enough
ionizing photons have been produced to ionize every hydrogen atom
exactly once.  The solid line is for model~A, the dotted for model~G
and the dashed for model~C.
\label{fig:nphnb}
}
\vspace{1cm}

An interesting issue is whether massive stars in young galaxies
can, on their own, produce enough ionizing photons ($\lambda<912{\rm
\AA}$) to re-ionize the IGM by $z=5$. (Additional ionizing photons are
produced by quasars, and possibly by pregalactic stars formed at high
redshift via molecular hydrogen cooling, e.g. Tegmark \etal 1997). A
very simple criterion for establishing if re-ionization is possible by
redshift $z$ is based on the ratio $n_{\gamma}/n_{\rm H}$, where
$n_{\gamma}$ is the total number of ionizing photons produced per
comoving volume up to redshift $z$, and $n_H$ is the total comoving
density of hydrogen atoms. When
$n_{\gamma}/n_{\rm H}=1$, just enough ionizing photons are produced to
ionize each hydrogen atom exactly once. This criterion neglects
absorption of photons within the emitting galaxies, recombination of
hydrogen in the IGM, and depletion of the IGM by collapse of gas into
dark halos. In particular, the fraction $f_{\rm esc}$ of ionizing photons
that escape from a galaxy may be quite low: Dove \& Shull (1994)
estimate $f_{\rm esc}\sim 10\%$ for our own galaxy, while Leitherer \etal
(1995) estimate $\langle f_{\rm esc}\rangle \lsim 3\%$ from far-UV
observations of nearby starburst galaxies. The photon number density, 
$n_{\gamma}$, is also sensitive to the form of the IMF above $10\msun$. 
Fig. 18 shows the dependence of $n_{\gamma}/n_{\rm H}$ 
on redshift for models A, C and G.

\section{Discussion} 
\label{s:conc}

We have used a semi-analytic model of galaxy formation to
interpret recent data on galaxy formation and evolution within the context
of hierarchical clustering theories, focussing primarily on the properties
of the recently discovered population of Lyman-break galaxies at $z\simeq
3$. Our modelling technique allows us to identify the role that this
population plays within the general scheme of galaxy formation, and to
relate these observations to other data at lower redshifts.  Our models are
quite general, but they inevitably require a number of assumptions and
simplifications most of which, in fact, reflect our poor understanding of
the processes of star formation and feedback.  Within these limitations we
attempt to represent the relevant physics using scaling laws that involve
the smallest possible number of free parameters. Our modelling strategy is
based on fixing these free parameters by requiring that the models should
match a small subset of the local data, particularly the field B-band
galaxy luminosity function. Thus specified, the models possess predictive
power and can be tested against high redshift data.

The specific cold dark matter models that we have considered all share the
feature that they reproduce the observed abundance of present day rich
clusters of galaxies. This fixes the amplitude of primordial density
fluctuations which determines the epoch at which structures on any mass
scale form.  Current large-scale structure data allow a range of values for
the cosmological parameters $\Omega_0$ and $\Lambda_0$.  By way of
illustration, we have explored in detail a critical density model and two
low-density models, one open and the other flat.  A second common feature
of the models we have considered is that they all agree at some level with
most data on the evolutionary properties of galaxies at relatively modest
redshifts, $z\lsim 1$, such as the evolution of the luminosity function
(Baugh \etal 1996a), and the counts of galaxies as a function of magnitude,
morphological type, and redshift (Cole \etal 1994, Baugh \etal 1996b, Frenk
\etal 1996). Several interesting predictions of the models have been
corroborated by subsequent data, including, for example, the redshift
distribution of $B=24$ counts (Frenk \etal 1997). 

An important prediction of semi-analytic models which also predated the
acquisition of the relevant data is the cosmic star formation history,
first discussed by White \& Frenk (1991) and calculated by Cole \etal
(1994) and Heyl \etal (1995) for the specific models discussed here (c.f.
Fig. 14). We showed in Section~5 that Madau's (1996) recent
data agree well with these model predictions, although uncertainties remain
in this comparison because of the unknown effects of dust obscuration and
possible incompleteness in the observational samples at high redshifts.  In
particular, the possibility that a significant fraction of the total star
formation may have been missed in recent surveys if it occured in dust
enshrouded starbursts at very high redshift has been the subject of some
recent debate. Such a population is not predicted in our models, although a
certain amount of dust obscuration at high redshift can be accommodated and
may even be required: our models, in fact, predict star formation rates
which are somewhat higher than those inferred from the data uncorrected for
dust obscuration (c.f. Fig.~16). The star formation rate in our models
peaks around $z=1-2$, but it never varies by more than an order of
magnitude over the entire range $0<z<6$. In fact, the star formation rate
at $z\simeq 5$ is almost identical to the star formation rate at the
present day. Nevertheless, half of all the stars present today only formed
since $z<1$.  The significance of the Lyman-break galaxies in the context
of these models lies in the fact that at the epoch when these galaxies are
observed, the process of star formation is just beginning in earnest. Thus
the Lyman-break galaxies are the first massive objects to sustain
appreciable star formation and, in this sense, they signal the onset of
galaxy formation.

The {\it distribution} of star formation rate at different redshifts
provides even stronger constraints on models than the evolution of the
integrated star formation rate. Our models match existing data quite well,
and further observational measurements of this fundamental quantity are
very important. In both models and observations, the quiescent star
formation rates are relatively low even in the largest protogalaxies, with
the majority of galaxies never forming stars at rates exceeding a few solar
masses per year. The exception to this are galaxies that undergo a burst of
star formation as a result of experiencing a major merger. However,
predictions for the strength, duration and frequency of such bursts will
require more detailed modelling that we have attempted so far.

In our models, the Lyman-break galaxies form from rare peaks in the
density field at high redshift. As a result, we predict that their spatial
distribution at $z\simeq 3$ should be strongly biased relative to the
underlying mass, with a typical bias parameter, $\bar{b} \simeq 4$, and a
comoving clustering length, $r_0\simeq 4 \mpc$. Generically, we 
expect the Lyman-break 
galaxies to be rotating disks, and a simple model for the
origin of their angular momentum predicts typical half-light radii of
$\simeq 0.5 \kpc$. The Lyman-break galaxies seen at $z\simeq 3$ evolve
into the present day ordinary ellipticals and spirals that make up the
bright end of the luminosity function.  
Their stars will typically be concentrated in the central regions of their 
descendants. These descendants are to be found
preferentially in groups and clusters, reflecting their biased origin
and strong clustering at high redshift.

The appearance of the first protogalaxies at $z\simeq 3.5$ and the late
conversion of most of the gas into stars fit in well with the observation
that the neutral hydrogen content of the universe, as determined from
measurements of damped Lyman-alpha clouds, peaks at around $z=3$ and
declines thereafter (Storrie-Lombardi \etal 1996). The neutral hydrogen
density in our models also exhibits this overall behaviour and agrees
reasonably well with the data once corrections for incompleteness and a
small amount of dust obscuration are included. Related semi-analytic models
by Kauffmann (1996b) agree even better with these data.  Our models are
consistent with Madau's (1996) view that, with the discovery of Lyman-break
galaxies, the bulk of the star formation (and the attendant metal
production) in our universe has, in effect, been identified. Only a small
fraction of the star formation activity remains to be detected during the
``dark ages'' prior to $z=4$. Characterising such activity is, of course,
of great importance for testing the general view that galaxies formed by
hierarchical clustering. According to our models, the small percentage of
stars that formed prior to the Lyman-break galaxy epoch produce enough
radiation to make at least a significant contribution to the UV flux
required photoionize the intergalactic medium by $z\simeq 5$.

Although a late epoch of galaxy formation in the standard CDM model was 
predicted long ago (Frenk \etal 1985, White \& Frenk 1991), a surprising
result of our analysis is that this is also true of the now popular
low-density variants of this model. Indeed, the star formation histories of
the $\Omega_0=1$ model and the flat $\Omega_0=0.3$ model 
are remarkably similar. This is largely coincidental:
the detailed star formation histories depend not only on the shape of the
power spectrum, but also on its normalisation and on the way in which star
formation and feedback are implemented in our galaxy formation models. The
main conclusion of our analysis is that, regardless of the exact values of
the cosmological parameters, CDM models that approximately reproduce the
abundance of Lyman-break galaxies, require massive galaxy formation to
begin around $z\simeq 3.5$.

In summary, we have argued in this paper that the main ingredients of a
consistent picture of galaxy formation may now be in place. The key
observation that has unlocked this paradigm is the discovery of a large
population of star-forming galaxies at $z\simeq 3$ which signal the onset
of the epoch of galaxy formation that extends well into the present day. At
this time, data and theoretical modelling paint only a broad brush picture
of how galaxy formation may have occurred. Fortunately, if this emerging
picture is correct, the details should also be accessible to current
observational and modelling capabilities.

\section*{Acknowledgements}
We thank Charles Steidel for supplying the transmission of the filters used
in his observations and for valuable discussions.  We would like to thank
the referee, Piero Madau, and Simon White for their careful reading of the
manuscript and for helpful comments which helped improve the final version
of this paper.  We also acknowledge discussions with Richard Ellis, Michael
Fall, Max Pettini and Martin Rees. This research  was supported 
by the European Commission through the TMR Network on `` The Formation 
and Evolution of Galaxies'' and in part by a PPARC rolling grant.  
SMC acknowledges a PPARC Advanced Fellowship and CSF
acknowledges a PPARC Senior Fellowship. CGL was supported by the Danish
National Research Foundation through its establishment of the Theoretical
Astrophysics Center. 

\section*{References}

\setlength{\parindent}{0cm}

\def\refe {\par \hangindent=1cm \hangafter=1 \noindent}
\def\Astron {Astron}
\def\Astroph {Astroph}
\def\aj { \Astron. J., \rm}
\def\apj { \Astroph. J., \rm }
\def\apjs { \Astroph. J. Suppl., \rm }
\def\mn { MNRAS, }
\def\apl { Ap. J. (Letters), }

\refe Abraham, R.G., Tanvir, N.R., Santiago, B.X., Ellis, R.S., 
      Glazebrook, K., Van den Bergh, S., 1996 \mn 279, L47 
\refe Baron, E., White, S.D.M., 1987, \apj 322, 585
\refe Baugh, C.M., Efstathiou, G., 1993, \mn 265, 145
\refe Baugh, C.M., Cole, S., Frenk, C.S., 1996a, \mn 282, L27
\refe Baugh, C.M., Cole, S., Frenk, C.S., 1996b, \mn 283, 1361
\refe Bennet, C.L. \and 1996, \apj 464 L1
\refe Bond, J.R., Cole, S., Efstathiou, G., Kaiser, N., 1991, \apj 379, 440
\refe Bower, R.G., 1991, \mn 248, 332
\refe Brainerd, T.G., Smail, I., Mould, J. 1995, \mn 275, 781
\refe Bruzual, G., Charlot, S., 1993, \apj 405, 538
\refe Charlot, S., Worthey, G., Bressan, A., 1996, \apj 457, 625
\refe Cole, S., 1991, \apj 367, 45
\refe Cole, S., Kaiser, N., 1988, \mn 233, 637
\refe Cole, S., Arag\'{o}n-Salamanca, A., Frenk, C.S., Navarro, J.F., Zepf, 
      S.E., 1994, \mn 271, 781
\refe Cole, S., Weinberg,  D.H.,  Frenk, C.S., Ratra, B., 1997, \mn 289, 37 
\refe Connolly, A.J., Szalay, A.S., Dickinson, M., SubbaRao, M.U., 
      Brunner, R.J. 1997, \apj 486, L11
\refe Copi, C.J., Schramm, D.N., Turner, M.S., 1996, Nucl. Phys. B., S51B, 66.
\refe Cowie, L.L., Songaila, A., Hu, E.M., Cohen, J.G., 1996,  
      \apj 112, 839
\refe Dove, J.B, Shull, J.M., 1994, \apj 430, 222
\refe Dressler, A., Oemler, A., Sparks, W.B., Lucas, R.A., 1994, \apj 435, L23 
\refe Driver, S.P., Windhorst, R.A., Griffiths, R.E., 1995, \apj\, 453, 48
\refe Eggen, O.J., Lynden-Bell, D., Sandage, A.R., 1962, \apj 136, 748
\refe Eke, V.R., Cole, S., Frenk, C.S., 1996, \mn 282, 263
\refe Ellis, R.S., Colless, M., Broadhurst, T., Heyl, J., 
      Glazebrook, K., 1996, MNRAS, 280, 235
\refe Frenk, C.S., Baugh, C.M., Cole, S., 1996, IAU Symposia 171, 247
\refe Frenk, C.S., Baugh, C.M., Cole, S., Lacey, C.G., 1997 to appear 
      in Dark Matter and Visible Matter in Galaxies,
      eds. M. Persic \& P. Salucci.
\refe Gallego, J., Zamorano, J., Arag\'{o}n-Salamanca, A., Rego, M.,
      1995, \apj 455, L1
\refe Giavalisco, M., Steidel, C.C., Macchetto, F.D., 1996, \apj 470, 189
\refe Glazebrook, K., Ellis, R., Santiago, B., Griffiths, R., 1995a, \mn 
      275, L19.
\refe Glazebrook, K., Ellis, R., Colless, M., Broadhurst, T., Allington-Smith, J., 
      Tanvir, N., 1995b, \mn 273, 157
\refe Guiderdoni, B., Hivon, E. Bouchet, F. R., Maffei, B. 1997a, \mn in press.
\refe Guiderdoni, B., Bouchet, F. R., Puget, J. L., Lagache, G, \& Hivon,
      E. 1997b, Nature, in press. 
\refe Heyl, J.S., Cole, S., Frenk, C.S., Navarro, J.F., 1995, \mn 274, 755
\refe Kashlinsky, A., Mather, J.C., Odenwald, S., 1996, \apj 437 L9. 
\refe Kauffmann, G., 1995, \mn 274, 161
\refe Kauffmann, G., 1996a, \mn 281, 475
\refe Kauffmann, G., 1996b, \mn 281, 487
\refe Kauffmann, G., White, S.D.M., Guiderdoni, B., 1993, \mn 264, 201
\refe Kauffmann, G., Guiderdoni, B., White, S.D.M., 1994, \mn 267, 981
\refe Kauffmann, G., Charlot, S., White, S.D.M., 1996 \mn 283, L117
\refe Lacey, C.G., Cole, S. 1993 \mn 262, 627
\refe Lacey, C.G., Guiderdoni, B., Rocca-Volmerange, B., Silk, J.,
1993, \apj 402, 15
\refe Lacey, C.G., Silk, J., 1991 \apj 381,14
\refe Lanzetta, K.M., Wolfe, A.M., Turnshek D.A., 1995, \apj 440, 435
\refe Leitherer, C., Ferguson, H.C., Heckman, T.M., Lowenthal, J.D.,
1995, \apj 454, L19
\refe Liddle, A. R., Lyth, D. H., Viana, P. T. P.,  White, M., 1996, \mn
282, 281
\refe Lilly, S.J., Tresse, L., Hammer, F., Crampton, D., LeFevre, O, 
       1995, \apj 455, 108
\refe Lilly, S.J., LeFevre, O, Hammer, F., Crampton, D., 
       1996, \apj 460, L1
\refe Limber, D. N., 1954, \apj 119, 655
\refe Loveday, J., Peterson, B.A., Efstathiou, G., Maddox, S.J., 1992, 
      \apj 390, 338
\refe Lowenthal, J. D., Koo, D. C., Guzman, R., Gallego, J., Phillips,
A. C., Faber, S. M., Vogt, N. P., Illingworth, G. D., Gronwall, C., 1997,
\apj 481, 673 
\refe Lu, L.M., Sargent, W.L.W., Womble, D.S., Takadahidai, M., 1996, 
      \apj 472, 509
\refe Marzke, R.O., Huchra, J.P., Geller, M.J., 1994, \apj 428, 43
\refe Madau, P., 1995, \apj 441, 18
\refe Madau, P., 1996, preprint astro-ph/9612157
\refe Madau, P., Ferguson, H.C, Dickinson, M., Giavalisco, M.,
Steidel, C.C., Fruchter, A., 1996, \mn 283, 1388
\refe Madau, P., Pozzetti, L., Dickinson, M., 1997, submitted to \apj 
      astro-ph/9708220
\refe Meurer, G.R., Heckman, T.M., Lehnert, M.D., Leitherer, C., Lowenthal, J.
      1997, \aj,  114, 54
\refe Miller, G.E., Scalo, J.M., 1979, \apjs 41, 513
\refe Mo, H.J., Fukugita, M., 1996, \apj 467, L9
\refe Mo, H.J., Jing, Y.P., White, S.D.M., 1996, \mn 282, 1096
\refe Mo, H.J., White, S.D.M., 1996, \mn 282, 347
\refe Navarro, J.F., Frenk, C.S., White, S.D.M., 1996, \apj 462, 563
\refe Odewahn, S.C., Windhorst, R.A., Driver, S.P., Keel, W.C., 
      1996, \apj 472, L13
\refe Pascarelle, S.M., Windhorst, R.A., Keel, W.C., Odewahn, S.C.
      1996, Nature, 383, 45
\refe Peacock, J.A., Dodds, S.J., 1996, \mn 280, L19
\refe Peebles, P.J.E., 1980, Large Scale Structure in the Universe 
      Princeton.
\refe Pei, Y.C., Fall, S.M, 1995, \apj 154, 69
\refe Pettini, M., Smith, L.J., King, D.L., Hunstead,R.W., 1997, 
      \apj 486, 665
\refe Pettini, M., Steidel, C.C., Adelberger, K.L., Kellogg, M., 
      Dickinson, M., Giavalisco, M., 1997, To appear in `ORIGINS', 
      ed. J.M. Shull, C.E. Woodward, and H. Thronson, 
      (ASP Conference Series) (astro-ph/9708117) 
\refe Press, W.H., Schechter, P.L., 1974, \apj 187, 425
\refe Puget, J.L., Abergel, A., Bernard, J.P., Boulanger, F., 
      Burton, W.B., Desert, F.X., Hartmann, D., 1996, Astron. Astrop., 
      308, L5 
\refe Sawicki, M.J., Lin, H., Yee, H.K.C. 1997, \aj 113, 1
\refe Salpeter, E.E., 1955, \apj 121, 61
\refe Scalo, J.M., 1986, Fundamentals of Cosmic Physics, 11, 1
\refe Smail, I., Dressler, A., Kneib, J.P., Ellis, R.S., Couch, W.J., 
      Sharples, R.M., Oemler, A., 1996, \apj, 469, 508.
\refe Smail, I., Dressler, A., Couch, W.J., Ellis, R.S.,  
      Oemler, A., Butcher, H., Sharples, R.M.,  1997, \apjs 110, 213
\refe Steidel, C.C., Hamilton, D., 1992, \aj 104, 941
\refe Steidel, C.C., Hamilton, D., 1993, \aj 105, 2017
\refe Steidel, C.C., Pettini, M., Hamilton, D., 1995, \aj 110, 2519
\refe Steidel, C.C., Giavalisco, M., Pettini, M., Dickinson, M., 
      Adelberger, K.L., 1996a, \apj 462, L17 (S96).
\refe Steidel, C.C., Giavalisco, M., Dickinson, M., Adelberger, K.L., 
      1996b, \aj 112, 352
\refe Storrie-Lombardi, L.J., McMahon, R.G., Irwin, M.J., 1996, \mn 283, L79
\refe Tegmark, M., Silk, J., Rees, M.J., Blanchard, A., Abel, T.,
Palla, F., 1997, \apj 474,1
\refe Treyer, M.A., Ellis, R.S., Milliard, B., Donas, J., 1997, to appear in  
      ``The Ultraviolet Universe at Low and High Redshift: Probing the 
        Progress of Galaxy Evolution", AIP press
\refe Tytler, D., Fan, X.M., Burles, S., 1996, Nature, 381, 207
\refe Viana, P.T.P., Liddle, A.R., 1996, \mn 281, 323 
\refe White, S.D.M., Efstathiou, G., Frenk, C.S., 1993, \mn 262, 1023
\refe White, S.D.M., Frenk, C.S., 1991, \apj 379, 52
\refe White, S.D.M., Rees, M.J., 1978, \mn 183, 341. 
\refe Williams, R.E., et al. 1996, \aj 112, 1335.
\refe Wolfe A.M., Lanzetta K.M., Foltz C.B., Chaffee, F.H., 1995, \apj 454, 698

\end{document}